\documentclass[journal]{IEEEtran}
\usepackage{cite}
\usepackage{multicol,lipsum}
\usepackage{amsmath,amssymb,amsfonts}
\usepackage{algorithmic}
\usepackage{graphicx}
\usepackage{textcomp}
\usepackage{xcolor}
\usepackage{booktabs}
 \usepackage{hyperref}
\usepackage{enumerate}
\usepackage{algorithm}
\usepackage{multirow}
\usepackage{standalone}
\usepackage{tikz}
\usetikzlibrary{positioning}
\usetikzlibrary{decorations.markings}
\usepackage{float}

  \usepackage{enumerate}
 \usepackage{caption}
\usepackage{subcaption}
 \usepackage{setspace}
\usepackage{verbatim}
   \usepackage{tabularx}
 \usepackage{makecell}

 \usepackage{hyperref}

\begin{document}
%
\title{Enhancing  Signal Space Diversity  for SCMA   Over Rayleigh Fading Channels}
%
%
%

\author{
 Qu Luo,   ~\IEEEmembership{Graduate Student Member,~IEEE,}
       Zilong Liu, ~\IEEEmembership{Senior Member,~IEEE,}
       Gaojie Chen, ~\IEEEmembership{Senior Member,~IEEE,} 
       and
       Pei Xiao, ~\IEEEmembership{Senior Member,~IEEE.}
 
\thanks{ Qu  Luo, Gaojie Chen and  Pei  Xiao   are  with  5G \& 6G  Innovation Centre, Institute for Communication Systems (ICS), University of Surrey, UK, email:\{q.u.luo,  gaojie.chen, p.xiao\}@surrey.ac.uk.}
\thanks{ Zilong   Liu   is   with   the   School   of   Computer   Science   and   Electronic   Engineering,   University   of   Essex,   UK,    email:   zilong.liu@essex.ac.uk.}
 }
\maketitle

\begin{abstract}
Sparse code multiple access (SCMA) is a promising technique for the enabling of massive connectivity in future machine-type communication networks, but it suffers from a limited diversity order which is a bottleneck for significant improvement of error performance.  This paper aims for enhancing the signal space diversity of sparse code multiple access (SCMA) by introducing quadrature component delay to the transmitted codeword of a downlink SCMA system in Rayleigh fading channels. Such a system is called SSD-SCMA throughout this work.  By looking into the average mutual information (AMI) and the pairwise error probability (PEP) of the proposed SSD-SCMA, we develop novel codebooks by maximizing the derived AMI lower bound and a modified minimum product distance
(MMPD), respectively. The intrinsic asymptotic  relationship between the AMI lower bound and   proposed MMPD based codebook designs is    revealed. Numerical results show   significant error performance improvement in the both uncoded and coded SSD-SCMA systems.

\end{abstract}


%
\IEEEpeerreviewmaketitle

\section{Introduction}

\IEEEPARstart{T}he widespread proliferation of wireless services and internet-of-thing (IoT) devices is challenging   the  legacy human-centric mobile  networks. For higher spectral efficiency and lower communication latency, there has been  a paradigm shift in recent years in the study of non-orthogonal multiple access (NOMA), where  the same time/frequency resources are shared for the supporting of several times of    more active users \cite{liu2021sparse,ReinforcementQu}. Among many others, this work    is concerned with a  representative code-domain NOMA (CD-NOMA) technique called   sparse code multiple access (SCMA), where multiple users communicate concurrently with distinctive  sparse codebooks  \cite{nikopour2013sparse}. 
At the transmitter,   the  incoming message bits of each SCMA user are directly mapped to a multi-dimensional sparse codeword drawn from a carefully designed codebook \cite{taherzadeh2014scma,NovelLuo}.   As pointed out in \cite{liu2021sparse}, the conventional  SCMA (C-SCMA) suffers from    a small diversity order which is a critical bottleneck for fundamental improvement of the SCMA error performance.  Therefore, it is pivotal  to look for new and affordable SCMA transmission schemes for significant enhancement of its system signal space diversity (SSD).

\subsection{Related Works}  

SSD, as an effective transmission scheme for higher diversity,  has received a sustained  research attention in the past decades. A power- and bandwidth-efficient way to acquire SSD was proposed in  \cite{BoutrosSignal} by coordinate interleaving with constellation rotation.  
For significant performance gain, the rotation angles of different constellations were investigated  in  \cite{khormuji2006rotation,5379002} for uncoded systems and in \cite{CodedXie, DesignChindapol} for  bit-interleaved coded modulation (BICM) systems.  In  \cite{IncioFull}, a different approach to attain full diversity was proposed by judiciously permuting a one-dimensional real-constellation through combinatorial optimization to form a multi-dimensional codebook. This was soon followed by \cite{IncioLow} with a low-complexity list-based detection algorithm that works for SSD with both partial- and full-diversity multi-dimensional codebooks.  It is noted  that the above works \cite{BoutrosSignal, khormuji2006rotation,5379002,CodedXie, DesignChindapol,IncioFull,IncioLow} were mainly conducted in a single user system with OMA transmission. As far as multiuser communication is concerned,   there have been some works on the exploiting the  SSD in power-domain NOMA (PD-NOMA) systems \cite{DownlinkQiu,zyurtQuadrature,PandeyCoordinated,ZhangPower,JointNg}. In these works, the composite constellations with different rotation angles  were obtained for two or more users by optimizing certain criteria such as minimum distance, maximum mutual information (MI), and minimum   pairwise error probability  (PEP). To the best of our knowledge, however, no results are known on SSD assisted  SCMA.

An important aspect of SCMA system optimization is sparse codebook design in order to achieve excellent error rate performances in different channel conditions.   Existing codebook  designs  mainly follow  a multi-stage design optimization by first constructing a common multidimensional constellation, called a mother constellation (MC), upon which certain user-specific operations (e.g., interleaving, permutation, shuffling and phase
rotations) are applied  to the MC to obtain codebooks for multiple users \cite{yu2015optimized,mheich2018design,chen2020design,Zhang,luo2023design}. In general, the   MC and user-specific operations can be designed by minimizing the PEP conditioned to certain channel conditions  \cite{yu2015optimized,mheich2018design,chen2020design,Zhang,luo2023design,huang2021downlink} or maximizing the system    capacity  \cite{dong2018efficient,BaoDesign,XiaoCapacity,Sharma,JiangLow}. 
By looking into the PEP  over Gaussian and Rayleigh fading channels, it is desirable to maximize  the minimum Euclidean
distance (MED) and  minimum product distance (MPD) of a MC or a codebook.  Following this spirit,    \cite{yu2015optimized} considered Star-QAM as the MC for enlarged   MED of   the superimposed codewords in downlink SCMA systems.    Golden angle modulation (GAM)  constellation was adopted in   \cite{mheich2018design}  to construct SCMA codebooks with   low  peak-to-average power ratio  properties.  In \cite{chen2020design}, near-optimal  codebooks for different channel conditions  were investigated by choosing suitable MCs with large MPD. A  uniquely decomposed constellation group   based codebook design approach was proposed in \cite{Zhang} by maximizing the MED at each resource node and the MPD of the MC. Downlink quaternary sparse codebook with large MED was obtained in \cite{huang2021downlink}  by solving a non-convex optimization problem. Recently,   a novel class of low-projection SCMA codebooks for     ultra-low decoding complexity was  developed in \cite{luo2023design} by maximizing  the proposed distance metric over Rician fading channels.

SCMA codebooks can also be optimized from the capacity perspective, as shown in  \cite{dong2018efficient,BaoDesign,XiaoCapacity,Sharma,JiangLow}.   
In \cite{dong2018efficient}, a gradient based algorithm was proposed to optimize the average mutual information (AMI), where the AMI was calculated by Monte Carlo method due to the unavailability of its closed form  \cite{dong2018efficient}. 
To avoid the prohibitively high-complexity AMI computation, the cutoff rate was considered in SCMA codebook optimization in   \cite{BaoDesign,Sharma,XiaoCapacity}. 
 Specifically,   \cite{BaoDesign} proposed  a performance criterion based on the cutoff rate of the equivalent multiple-input multiple-output SCMA system for uplink  Rayleigh fading channels.  In \cite{XiaoCapacity}, new MCs   were obtained by looking into the constellation constrained sum rate capacity.   The cut-off rate combined with constellation  shaping gain  were  considered  in \cite{Sharma}.  More recently, a novel  sparse codebook was    obtained    in  \cite{JiangLow}  by maximizing the derived lower bound of AMI.  However, the  lower bound with closed-form of AMI for Rayleigh fading channels is still  missing.  It is noted that the $M$-order pulse-amplitude modulation ($M$-PAM) was    employed as the basic constellation in \cite{XiaoCapacity,Sharma,JiangLow}, thus their  resultant codebooks  exhibit    certain similarity.  



\subsection{Motivations and Contributions }

Against the aforementioned background, the  motivations of this work are the two-fold: 1) As  SSD can provide enhanced diversity gain over fading channels,  a fundamental investigation on the  amalgamation of SSD and SCMA, refereed to as  SSD-SCMA, is necessary on the theoretical trade-offs and design guidelines;   2)  Albeit there are  numerous SCMA codebook designs based on PEP or capacity, these codebooks  may not be optimal for   SSD-SCMA. 


The main novelties and contributions of the paper are summarized as follows:
 
 
 
 \begin{itemize}
     \item We   introduce quadrature component delay to the superimposed codeword of a downlink SCMA for efficient  acquisition of SSD, where the resultant system is called SSD-SCMA. Interestingly, we show that the resultant diversity order is doubled compared to that in conventional SCMA, thus leading to significantly improved  reliability in Rayleigh fading channels.  To guide the system design,   an AMI   lower bound   and a  PEP upper bound  are derived.

     \item Based on the derived AMI lower bound and the modified minimum product distance (MMPD) from the proposed PEP upper bound, we     formulate  systematic  design metrics including the MC design, sparse codebook optimization, and bit  labeling  from both the PEP and AMI perspectives.    In addition, we fill a gap in the current SCMA literature on the  asymptotic relationship between PEP and  AMI based design metrics, thus bridging  the fundamental connection of these two   SCMA codebook design techniques.    
     

   \item We develop  an   enhanced GAM (E-GAM)   as the $N$-dimensional MC for the proposed  AMI based codebooks (AMI-CBs).  
   The joint optimization of MC and rotation angles by maximizing the AMI  lower bound  are carried out   with an interior point method (IPM) with random initial values and Monte Carlo sample estimation.  For the proposed PEP based codebooks (P-CBs), we advocate the permutation of a  basic one-dimensional constellations that owns large MED to construct the $N$-dimensional MC.  The rotation angles for different  users are optimized based   on the proposed multi-stage   search.
    
    \item We conduct extensive numerical experiments  to show the superiority of the proposed SSD-SCMA systems and the proposed codebooks in both uncoded and  BICM with    iterative demapping and decoding (BICM-IDD) systems.  The simulations  indicate  that significant error performance gains are  achieved for SSD-SCMA with the proposed AMI-CBs and P-CBs compared to the C-SCMA systems with the state-of-the-art codebooks.

 \end{itemize}
 
     \subsection{Organization}

The rest of the paper is organized as follows. In Section II, the system model of downlink SSD-SCMA along with  the  multiuser detection technique are presented.   Section III analyzes the AMI and PEP of the SSD-SCMA system in Rayleigh fading channels. In Section IV, the  codebook design   problems  for SSD-SCMA are formulated in terms of the AMI and PEP. The detailed design of AMI-CB and P-CB is elaborated in Section V.  The numerical results are given in Section VI. Finally, conclusions are made in Section VII.
           
           \subsection{Notation}
  The $n$-dimensional complex, real and binary vector spaces are denoted as $\mathbb{C}^n$, $\mathbb{R}^n$ and $\mathbb{B}^n$, respectively.  Similarly, $\mathbb{C}^{k\times n}$, $\mathbb{R}^{k\times n}$ and $\mathbb{B}^{k\times n}$  denote the $(k\times n)$-dimensional complex, real and binary  matrix spaces, respectively. ${{\mathbf{I}}_{n}}$ denotes an $n \times n $-dimensional  identity matrix. $\text{tr}(\mathbf{X})$ denotes the trace of a square matrix $\mathbf{X}$.   $\text{diag}(\mathbf{x})$ gives a diagonal matrix with the diagonal vector of $\mathbf{x}$. $(\cdot)^\mathcal T$, $(\cdot)^ \dag $ and $(\cdot)^\mathcal H$ denote the transpose, the conjugate and the Hermitian transpose operation, respectively.  $\|\mathbf{x}\|_2$ and $|x|$ return the Euclidean norm of vector $\mathbf{x}$ and the absolute value of $x$, respectively.  $\mathbf{x}_{\mathrm I}$   and $\mathbf{x}_{\mathrm Q}$  return the in-phase ($\mathrm I$) and quadrature ($\mathrm Q$) components of the vector, respectively.

 \section{Introduction to the  proposed SSD-SCMA}
 
\subsection{Introduction to SCMA}
\begin{figure} 
  \centering
  \includegraphics[width=3in]{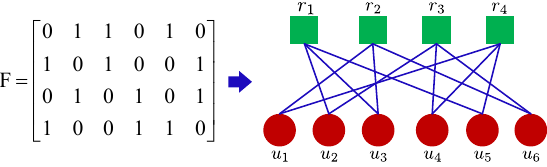} 
  \caption{Factor representation of a $(4\times 6)$  SCMA system.}
\label{factor}
\vspace{-0.5cm}
\end{figure}

We consider a   downlink  SCMA system where $J$ users communicate over $K$ orthogonal resources. The overloading factor, defined     as $ \lambda = \frac{J}{K} $, is   larger than $100\%$. On  the transmitter side,  each user   maps $\log_2\left(M\right)$   binary bits to a length-$K$ codeword $ \mathbf {x} _{j}$ drawn from a pre-defined codebook $ \boldsymbol {\mathcal {X}}_{j}  \in \mathbb {C}^{K \times M}$, where $M$ denotes  the modulation order. The mapping relationship is expressed as $ f_j:\mathbb{B}^{\log_2M \times 1}  \rightarrow { \boldsymbol {\mathcal X}}_{j}   \in \mathbb {C}^{K \times M}, {~\text {i.e., }}\mathbf {x}_{j}=f_{j}(\mathbf {b}_{j}) $,  where the codeword set for the $j$th user is given by $ \boldsymbol {\mathcal {X}}_{j}=\{\mathbf{x}_{j,1}, \mathbf{x}_{j,2},\ldots,\mathbf{x}_{j,M}\} $  and  $\mathbf {b}_{j}=[b_{j,1},b_{j,2},\ldots,b_{j,\log _{2} M}]^{\mathcal T}\in \mathbb{B}^{\log _{2} M \times 1}  $ stands for the $j$th user's instantaneous input binary message vector. 
The $K$-dimensional complex codewords in the SCMA codebook are sparse vectors with $N$ non-zero elements and $N < K$. The sparsity of the codebooks enables the low complexity message passing algorithm (MPA) detection  at receiver. 
Let $\mathbf{c}_{j}$ be a length-$N$ vector drawn from  $ \boldsymbol{ {\mathcal C}}_{j}\subset \mathbb {C}^{N \times M }$, where $   \boldsymbol{{\mathcal C}}_{j}$ is obtained  by removing all the zero elements  in  $ \boldsymbol{{ \mathcal X}}_{j}$.  We further define the mapping from $\mathbb{B}^{\log_2M}$ to  $ \boldsymbol{{\mathcal C}}_{j}$ as
$
g_{j}:\mathbb {B}^{\log _{2}M\times 1}\mapsto  \boldsymbol{{\mathcal C}}_{j}, \quad {~\text {i.e., }}\mathbf {c}_{j}=g_{j}(\mathbf {b}_{j}) 
$. The SCMA mapping   now can be re-written as 
\begin{equation} 
\label{scmaMapping}
f_{j}:\equiv \mathbf {V}_{j}g_{j}, \quad {~\text {i.e., }}\mathbf {x}_{j}=\mathbf {V}_{j}g_{j}(\mathbf {b}_{j}),
\end{equation}
where $\mathbf {V}_{j} \in \mathbb {B}^{K \times N} $ is a  mapping   matrix that maps the $N$-dimensional vector  to a $K$-dimensional sparse   codewords. The sparse structure of the $J$ SCMA codebooks can be represented by the indicator  matrix (factor graph) $\mathbf {F}_{K \times J} = \left [ \mathbf {f}_1, \ldots, \mathbf {f}_J \right] \subset \mathbb {B}^{K\times J}$ where  $\mathbf {f}_j = \text {diag}(\mathbf {V}_j\mathbf {V}_j^{\mathcal T})$.  An element of  ${\bf{F}}$ is defined as ${f_{k,j}}$ which takes the value of $1$ if and only if  the
user node  $u_j$ is connected to   resource node  $r_k$ and 0 otherwise.  Fig. \ref{factor}  illustrates an SCMA  factor graph with $J=6$, $K=4$ and $N=2$.  

\subsection{Proposed SSD-SCMA}

\begin{figure}[t]
  \centering
  \includegraphics[width=3.5in]{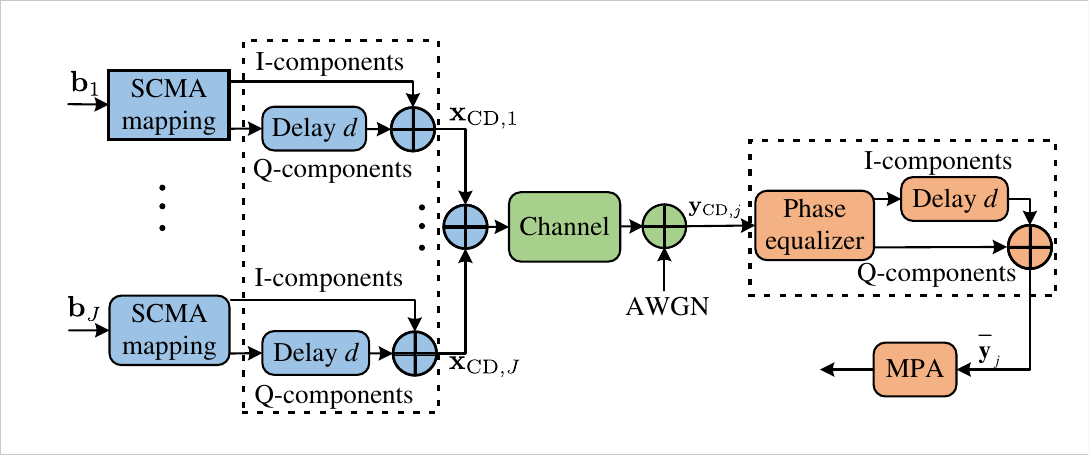} 
  \caption{Block diagram for the proposed SSD-SCMA system with quadrature component delay in a downlink Rayleigh fading channel. }
\label{SysMode}
\end{figure}

The key idea of SSD-SCMA is to introduce a delay $d$ for the quadrature component of the   SCMA codeword, where the delay time $d$ is assumed to be larger than the channel coherence time \cite{CodedXie, DesignChindapol}.   After the component delay (CD), the transmit signal of  $j$th user  is represented by  $\mathbf { x}_{\text{CD},j} $.  The block diagram for the proposed  SCMA systems with CD is shown in Fig. \ref{SysMode}.


 After the CD module, the transmitted vector is obtained by $\mathbf r_{\text{CD}}= \sum\nolimits_{j=1}^{J}{{{\mathbf{ x}}_{\text{CD},j}}}$. Accordingly, the received signal at the $j$th user can be written as
 \begin{equation}  \label{downlink}
\mathbf{  y}_{\text{CD},j}=\text{diag}\left( {\mathbf{ h}_{\text{CD},
j}} \right)\mathbf r_{\text{CD}}+\mathbf{z}_j,
\end{equation} 
where ${{\mathbf{h}}_{\text{CD},j}} \in {{\mathbb{C}}^{K\times 1}}$ is the  channel coefficient vector between  the base station and  the $j$th user,  and
${\mathbf{z}_j} \in {{\mathbb{C}}^{K\times 1}}$ is the complex additive white Gaussian noise (AWGN) vector  with the variance with zero mean and variance $N_{0}$. 
 We assume that perfect CSI is available at the receiver. After  the phase equalizer, the received signal is transformed into  $\frac{ \text{diag}\left( {{\mathbf{h}_{\text{CD},j}^\dag }}\right)  }{\mathrm{|}\text{diag}\left( {{\mathbf{h}_{\text{CD},j}} }\right) \mathrm{|}}\mathbf{ y}_{\text{CD},j} = \mathrm{|}\text{diag}\left( {{\mathbf{h}_{\text{CD},j}} }\right) \mathrm{|}\mathbf r_{\text{CD}} + \frac{ \text{diag}\left( {{\mathbf{h}_{\text{CD},j}^\dag }}\right)  }{\mathrm{|}\text{diag}\left( {{\mathbf{h}_{\text{CD},j}} }\right) \mathrm{|}}\mathbf{z}_j $   \cite{BoutrosSignal, khormuji2006rotation,5379002,CodedXie, DesignChindapol,IncioFull,IncioLow}. Since the noise  $\mathbf{z}_j$ is circularly  symmetric, $ \frac{ \text{diag}\left( {{\mathbf{h}_{\text{CD},j}^\dag }}\right)  }{\mathrm{|}\text{diag}\left( {{\mathbf{h}_{\text{CD},j}} }\right) \mathrm{|}}\mathbf{z}_j $ has the same distribution of noise $\mathbf{z}_j$.  
     Denote ${\overline {\mathbf{y}}}_{j}$  by the received signal after delaying the in-phase  component of the received $K$-dimensional vector. We further let  ${{{\mathbf{h}}_{j}^{\mathrm I}} } = \left[ h_{j,1}^{\mathrm I},h_{j,2}^{\mathrm I}, \ldots,h_{j,K}^{\mathrm I} \right] ^{\mathcal T}$ and  ${{{\mathbf{h}}_{j}^{\mathrm Q}} } = \left[ h_{j,1}^{\mathrm Q},h_{j,2}^{\mathrm Q}, \ldots,h_{j,K}^{\mathrm Q} \right]^{\mathcal T}$ be the channel gains  associated with  the $\mathrm I$ and $\mathrm Q$ components of the transmitted vector $\mathbf r$, respectively.  The elements of ${{{\mathbf{h}}_{j}^{\mathrm I}} }$ and ${{{\mathbf{h}}_{j}^{\mathrm Q}} }$ are Rayleigh distributed independent random variables with  zero mean and unit
variance. 
  Then, $\overline{\mathbf{y}}_{j}$ 
 can be demultiplexed into two independent parallel channels \cite{YaoDesign}:
 \begin{equation}
 \label{symodle}
 \begin{aligned}
  {\mathbf {y}}_j =   {\mathbf {H}_j} \mathbf{w}
 +\mathbf {n}_j  
 =  
 \left [{ \begin{array}{l}    \left |\text{diag}\left( {{{\mathbf{h}}_{j}^{\mathrm I}} }\right) \right |  \quad 0\\ 0\quad       \left | \text{diag}\left( {{{\mathbf{h}}_{j}^{\mathrm Q}} }\right) \right |     \end{array}}\right]
 \Bigg [{ \begin{array}{l}  {\mathbf{ {r} }}_\mathrm{I}\\  {\mathbf{ {r} }}_{\mathrm{Q}} \end{array}}\Bigg]\!+\!\Bigg [{ \begin{array}{l} \overline{\mathbf {z}} _{j, \mathrm I}\\ \overline {\mathbf {z}}_{j, \mathrm Q} \end{array}}\Bigg], 
 \end{aligned}
 \end{equation}
 where
 $    {\mathbf {y}}_j=[ {\overline {\mathbf y}}_{j,\mathrm I}^{\mathcal T}, {\overline {\mathbf y}}_{,j,\mathrm Q}^{\mathcal T}]^{\mathcal T} \in \mathbb R^{2K \times 1}$, $ {\mathbf{ {r} }} = \sum\nolimits_{j=1}^{J}{{{\mathbf{  x}}_{j}}}$,   and $\mathbf{w} = \left [  {\mathbf{ {r} }}_\mathrm{I}^{\mathcal T},    {\mathbf{ {r} }}_\mathrm{Q}^{\mathcal T}   \right ]^{\mathcal T} $.      $\mathbf{n}_j = \left [   \overline{\mathbf {z}} _{\mathrm I}^{\mathcal T},   \overline{\mathbf {z}}_\mathrm{Q}^{\mathcal T}   \right ]^{\mathcal T} $   is the real  Gaussian noise  vector  with the variance with zero mean and variance $\frac{N_{0} }{2}$, and  $\overline{\mathbf{z}}_j =   \frac{ \text{diag}\left( {{\mathbf{h}_j^\dag }}\right)  }{\mathrm{|}\text{diag}\left( {{\mathbf{h}_j} }\right) \mathrm{|}}\mathbf{z}_j$.  For simplicity, the  subscript $j$ in (\ref{symodle}) is omitted whenever no ambiguity arises.

Observed from (\ref{symodle}),  the $\mathrm{I}$ and $\mathrm{Q}$ components of the transmitted codewords  experience independent Rayleigh  fading channels.  We will show later that   the proposed SSD-SCMA along with efficient codebook design can significantly improve    the communication reliability in fading channels.

  \subsection{MPA Detection}
 The received signal $\overline {\mathbf y}$ will  be inputted into the      MPA decoder for efficient multi-user detection.   The MPA detector exploits the connections between the user nodes and resource nodes, and passes the belief information alongside the edges of the factor graph.  Define the sets $\varphi _j=\lbrace k:  {f}_{j,k} = 1 \rbrace$, representing all the  resource nodes that  user $j$ has active transmission, and $\phi _k=\lbrace j:  {f}_{j,k} = 1 \rbrace$, consisting of all the  users colliding over  resource node $k$. Following the basic principle  of the MPA,  at the $t$th iteration, the belief  message   propagating from   resource node ${r}_{k}$ to user node ${u}_{j}$, denoted by  $ I_{r_k \rightarrow u_j}^{t}(\mathbf {x}_j)$,  and the   belief  message   propagating from  user node ${u}_{j}$  to resource node ${r}_{k}$, denoted by $ I_{u_j \rightarrow r_k}^{(t)}(\mathbf {x}_j)$, can be expressed respectively as \cite{LuoError}
 \begin{equation}
  \small
  \label{FN_up}
 I_{r_k \rightarrow u_j}^{t}(\mathbf {x}_j)   = \sum _{\substack{ i\in \phi _k\backslash \lbrace j\rbrace \\ \mathbf {x}_i \in \mathcal {X}_{i}}}  p\left( \overline{y}_k \vert \mathbf {x}_i    \right) \prod _{i\in  \phi _k \backslash \lbrace j\rbrace } I_{u_i \rightarrow r_k}^{(t -1)}(\mathbf {x}_i), 
\end{equation}
and
\begin{equation}
\small
  \label{VN_up}
I_{u_j \rightarrow r_k}^{(t)}(\mathbf {x}) =\alpha _j  \prod _{\ell \in \varphi _j \backslash \lbrace k\rbrace } I_{r_{\ell } \rightarrow u_j}^{(t)}(\mathbf {x}), \end{equation}
where   $\overline{y}_k $ is the $k$th entry of $\overline{\mathbf  y} $,    $\alpha _j$ is  a normalization factor and the probability distribution function of $p\left( \overline {y}_k \vert \mathbf {x}_i    \right)$ is given by
\begin{equation}
\small
p\left( \overline{y}_k \vert \mathbf {x}_i    \right) = \frac{1}{\sqrt{2\pi N_{0}}} \text{exp} \left(-   \frac{  \sum _{l\in \{\mathrm{I}, \mathrm{Q}\}} \Big \vert \overline{y}_{l,k} -     \vert  h_{k}^{l}  \vert  \sum  _{i\in  \phi _k    } x_{l,k,i} \Big \vert^{2}}{2 N_0} \right).
\end{equation}

 \section{AMI Derivation and Error Performance Analysis}

This section  first derives  the  AMI and its lower bound of the proposed SSD-SCMA system, followed by the  error performance analysis based on the PEP. 
\subsection{The AMI of the Proposed SSD-SCMA}
\label{MI_sec}


Let    $\boldsymbol {\mathcal X} = \{\boldsymbol {\mathcal X}_1,\boldsymbol {\mathcal X}_2,\ldots,\boldsymbol {\mathcal X}_J\}$ denote the $J$ users' sparse codebooks. For the input vector $\mathbf w$    given in (\ref{symodle}),  the  AMI  of SSD-SCMA    is given by \cite{dong2018efficient}
\begin{equation}
\label{AMI}
\begin{aligned}
 \small
 \mathcal  I_{AMI} ^{\boldsymbol {\mathcal X}}
 &= \mathcal H \left( \mathbf{w} \right) - \mathcal H \left( \mathbf{w} \vert \mathbf y, \mathbf H\right)  \\   
 &  =J\log_2  (M) -{\mathbb{E}}_{{\mathbf w},{\mathbf y},\mathbf H}\left\{\log_{2}{\sum_{\hat{\mathbf w}\neq {\mathbf w}}p({\mathbf y}\vert \hat{\mathbf w},\mathbf H)\over p({\mathbf y}\vert {\mathbf w},\mathbf H)}\right\} \\
& =J\log_2  (M)-\frac{1}{{{M}^{J}}}\sum\limits_{m=1}^{{{M}^{J}}}{{{\mathbb E}_{ \mathbf H, \mathbf{n}} }\left\{\log \sum\limits_{p=1}^{{{M}^{J}}}{\exp \left( -{{d}_{m,p}} \right)} \right\}},
\end{aligned}
\end{equation}
where $ \mathcal H \left( \mathbf{w} \vert \mathbf y, \mathbf H\right) $ denotes   the entropy of $\mathbf w$ conditioned on $\mathbf H$ and  $\mathbf y$, and
 \begin{equation}
 \small
 \label{dmp}
     {{d}_{m,p}}=\frac{{{\left\| \mathbf{H}\left( {{\mathbf{w}}_{p}}-{{\mathbf{w}}_{m}} \right)+\mathbf{n} \right\|}^{2}}-{{\left\| \mathbf{n} \right\|}^{2}}}{{N_{0}}}.
 \end{equation}
 
  The AMI bounds the maximal information rate of the codebook set  $\boldsymbol {\mathcal X} = \{\boldsymbol {\mathcal X}_1,\boldsymbol {\mathcal X}_2,\ldots,\boldsymbol {\mathcal X}_J\}$ that can be reliably transmitted  with equiprobable  inputs.  In general, it is   challenging to obtain the AMI closed form. As an alternative solution, we deduce the    analytical lower   bound     of AMI to evaluate the transmit efficiency with finite  input   $\boldsymbol{\mathcal X}$.

\textbf{Lemma 1:} The AMI of the SSD-SCMA system in downlink Rayleigh fading channels is upper bounded by
  \begin{equation}
   \small
 \label{AMI_Up}
     \mathcal I_{UP}^{\boldsymbol {\mathcal X}}=  J\log(M)-    \sum\limits_{m=1}^{{{M}^{J}}}\log \left( { \sum\limits_{p=1}^{{{M}^{J}}}{\exp \left( -\frac{{{\left\|   {{\mathbf{r}}_{p}}-{{\mathbf{r}}_{m}}   \right\|}^{2}}  }{{N_{0}}} \right)}  } \right). 
 \end{equation}
 
 \textit{Proof:}   For   given $\mathbf H$, we have  
 \begin{equation}
 \label{AMI_up}
  \small
  \begin{aligned}
 \mathcal H \left( \mathbf{w} \vert \mathbf y, \mathbf H\right) 
    & \overset{(\mathrm i)}{\geq}    \sum\limits_{m=1}^{{{M}^{J}}}\log   \left ({ \sum\limits_{p=1}^{{{M}^{J}}}   \exp \left( {{\mathbb E}_{\mathbf{n}}} \left\{ -d_{m,p} \right) \right\}} \right ) \\
  & =  \sum\limits_{m=1}^{{{M}^{J}}}\log   \left ({ \sum\limits_{p=1}^{{{M}^{J}}}   \exp \left( - \frac{{{\left\| \mathbf{H}\left( {{\mathbf{w}}_{p}}-{{\mathbf{w}}_{m}} \right) \right\|}^{2}}}{{N_{0}}} \right) } \right ),\\  
      \end{aligned}
 \end{equation}
 where (i) is obtained by applying Jensen's inequality since the log-sum-exp function is a convex function of $d_{m,p}$.
Upon taking expectation of $\mathbf H$ on both sides of   (\ref{AMI_up}), we have 
\begin{equation}
 \label{AMI_up2}
  \small
  \begin{aligned}
 &\mathcal H \left( \mathbf{w} \vert \mathbf y \right)   = \mathbb E_{\mathbf H} \left\{ \mathcal H \left( \mathbf{w} \vert \mathbf y, \mathbf H\right)  \right\}   \\
   &  \overset{(\mathrm{ ii})}{\geq}  
      \sum\limits_{m=1}^{{{M}^{J}}}\log   \left ({ \sum\limits_{p=1}^{{{M}^{J}}}   \exp \left( \mathbb E_{\mathbf H} \left\{  - \frac{{{\left\| \mathbf{H}\left( {{\mathbf{w}}_{p}}-{{\mathbf{w}}_{m}} \right) \right\|}^{2}}}{{N_{0}}}  \right\} \right) } \right ) \\    &=     \sum\limits_{m=1}^{{{M}^{J}}}\log   \left ({ \sum\limits_{p=1}^{{{M}^{J}}}   \exp \left(  -   \frac{{{\left\|  \left( {{\mathbf{r}}_{p}}-{{\mathbf{r}}_{m}} \right) \right\|}^{2}}}{{N_{0}}}   \right) } \right ).
      \end{aligned}
 \end{equation}
 Substituting (\ref{AMI_up2}) into
 \begin{equation}
\label{MI}
 \small
 \mathcal {I}(\mathbf {w};\mathbf {y})  =J\log_2  (M)- \mathbb E_{\mathbf H} \left\{  \mathcal H \left( \mathbf{w} \vert \mathbf y, \mathbf H\right) \right\}     
\end{equation}
 yields the upper bound in (\ref{AMI_Up}).

\textbf{Lemma 2:} The AMI of the SSD-SCMA system in downlink Rayleigh fading channels is lower bounded by
  \begin{equation}
 \label{AMI_LR}
 \small
 \begin{aligned} 
   \mathcal  I_{LB}^{\boldsymbol {\mathcal X}} =   2J\log(M)-&   K \left( \frac{1}{\ln2}- 1 \right)    -  \log \left(  \sum\limits_{m=1}^{{{M}^{J}}}{ \sum\limits_{p=1}^{{{M}^{J}}}{\prod _{k=1}^{K}     \gamma _{k,m,p}}} \right) ,
     \end{aligned}
 \end{equation}
 where
\begin{equation}
\small
    {  \gamma _{k,m,p}= \prod _{l \in \{ \mathrm I, \mathrm Q\}}  \left ({ 1 + \frac {  \Big \vert  \sum \limits_{j \in \phi _k  }  x_{j,m,l}[k] - x_{j,p,l}[k] \Big \vert^{2}}{4N_{0}} }\right )^{-1}}.
\end{equation} 
 

 \textit{Proof:} By taking the noise term $\frac{ {{\left\| \mathbf{n} \right\|}^{2}}}{{N_{0}}}$ in ${{d}_{m,p}}$ out the summation, we can reformulate the $\mathcal H \left( \mathbf{w} \vert \mathbf y, \mathbf H\right)$ in (\ref{AMI}) as \cite{LowYang}
  \begin{equation}
 \label{AMI_re}
  \small
  \begin{aligned}
  &\mathcal H   \left( \mathbf{w} \vert \mathbf y, \mathbf H\right)  \\
    &=   {\rm E}_{{\bf n}}\log  \exp(\frac{\Vert {\bf n}\Vert^{2}}{N_{0}})      
     +\frac{1}{{{M}^{J}}}\sum\limits_{m=1}^{{{M}^{J}}}{{{\mathbb E}_{\mathbf{n}}}\left\{ \log \sum\limits_{p=1}^{{{M}^{J}}} \exp \left( e_{m,p} \right) \right\}}\\
   &  \overset{(\mathrm i)}{\leq}    \frac{K}{\ln2} 
     +   \log   \left ( \frac{1}{{{M}^{J}}} \sum\limits_{m=1}^{{{M}^{J}}}{ \sum\limits_{p=1}^{{{M}^{J}}} {{\mathbb E}_{\mathbf{n}}} \left\{   \exp \left( e_{m,p} \right) \right\}} \right ),
      \end{aligned}
 \end{equation}
where $e_{m,p}= -\frac{{{\left\| \mathbf{H} \left( {{\mathbf{w}}_{p}}-{{\mathbf{w}}_{m}} \right)+\mathbf{n} \right\|}^{2}}}{{N_{0}}}$. Considering the integral interval  of  $\left ( -\infty, \infty \right )$  for   the second term of the right-hand side, we have ${\rm E}_{{\bf n}} \left\{\log\exp(\Vert {\bf n}\Vert^{2}/N_{0})\right\}= \frac{K}{\ln2} $. Since $\log \left(x \right)$ is a concave  function,   an lower bound for the AMI in (\ref{AMI_re}) is derived  by applying  Jensen’s inequality, i.e., step (i).  The expectation over $\mathbf n $ in  (\ref{AMI_re}) is given by 
  \begin{equation}
   \small
\begin{aligned}
\label{cmk}
&   {{\mathbb E}_{\mathbf{n}}} \left\{  \exp   \left( e_{m,p} \right) \right\} \\
& =\int \frac {1}{\left ({\pi N_{0}}\right)^{K}}\exp   \Big(\frac {-\Vert{\mathbf n}\Vert^{2}}{N_0}\Big)      \exp   \left( e_{m,p} \right) {\mathrm {d}}{\mathbf  n} \\
 & =  \prod_{k=1}^{2K}  \frac {1}{\left ({\pi N_0}\right)^{K}}    \int_{n_k} \!\exp \Big( \! -\! \frac{{{\vert {n_k} +  \vert {h}_k \vert\left( {{{w}}_{p,k}}-{{{w}}_{m,k}} \right) \vert}^{2}  \!+ \!{\vert{ n_k}\vert ^{2}}}}{{N_0}}  \Big) {\mathrm {d}}{n_k}\\
& \overset{(\mathrm i)}{= }\frac {1}{2^{K}} \prod _{k=1}^{2K}   \exp \left ({ -\frac { \vert  {h}_{k} \vert^{2} \delta _{p,m,k}^{2}}{2N_{0}} }\right),
 \end{aligned}
 \end{equation}
where step (i) is derived based on the  (2.33.1)  in \cite{gradshteyn2014table},  $\delta_{p,m,k}^{2} = \left({{{w}}_{p,k}}-{{{w}}_{m,k}}\right)^2 $, $\vert h_k \vert $ is the $k$th entry of $ \text{diag}(\mathbf H)$, and  ${w}_{m,k} $ is the  $k$th  entry  of $\mathbf{w}_{m}$.  Then,    substituting   (\ref{cmk})  into (\ref{AMI_re}) and taking    expectation of $\mathbf H$ on the both sides,   we obtain
  \begin{equation}
 \label{AMI_lb1}
  \small
  \begin{aligned}
  &  \mathcal H   \left( \mathbf{w} \vert \mathbf y \right) = \mathbb E_{\mathbf H}  \left\{\mathcal H   \left( \mathbf{w} \vert \mathbf y, \mathbf H\right) \right\}   \\
  & \leq    \frac{K}{\ln2}-J\log(M)    \\ &    +  
      \mathbb E_{\mathbf{H}} \left\{\log   \left ( \sum\limits_{m=1}^{{{M}^{J}}}{ \sum\limits_{p=1}^{{{M}^{J}}}  \frac {1}{2^{K}} \prod _{k=1}^{2K} \exp \left ({ -\frac { \vert  {h}_{k} \vert^{2} \delta _{p,m,k}^{2}}{2N_{0}} }\right)} \right ) \right\}\\
   &  \ \leq  K \left( \frac{1}{\ln2}- 1 \right)- J\log(M) 
   \\ &    +   \log   \Bigg(  \sum\limits_{m=1}^{{{M}^{J}}}   \sum\limits_{p=1}^{{{M}^{J}}}     \prod _{k=1}^{2K}  \mathbb E_{\mathbf{H}} \left\{\exp \left ({ -\frac {  \vert  {h}_{k} \vert^{2} \delta _{p,m,k}^{2}}{4N_{0}} }\right)\right\}  \Bigg) \\
   & \overset{(\mathrm {i})} = K \left( \frac{1}{\ln2}- 1 \right)-J\log(M)  \\ &   +   \log \Bigg(  \sum\limits_{m=1}^{{{M}^{J}}} \sum\limits_{p=1}^{{{M}^{J}}}  \prod _{k=1}^{K} \prod _{l \in \{ \mathrm I, \mathrm Q\}}  \Bigg ({ 1 \!+ \! \frac {  \Big \vert  \sum \limits_{j  \phi _k }  x_{j,m,l}[k] - x_{j,p,l}[k] \Big \vert^{2}}{4N_{0}} }\Bigg )^{-1} \Bigg ) ,
      \end{aligned}
 \end{equation}
where step (i) is obtained base on the fact that the $s =h_k^2$ has a chi-square probability distribution with its moment generating function, which is defined as $\mathbb E \left[ e^{-st}\right]$, given by $M_s(t) = \frac{1}{1+t}$.
Substituting (\ref{AMI_lb1})  into (\ref{MI}) leads to the lower bound in (\ref{AMI_LR}).

\textit{Remark 1:} Following a similar derivation to the above for Lemma 2, we can obtain the lower bound of AMI for conventional SCMA,  which has the same form from (\ref{AMI_LR}), but with different expression of $\gamma _{k,m,p}$ given by
\begin{equation}
\small
\label{CSCMAMI}
    {  \gamma _{k,m,p}=  \left ({ 1 + \frac {  \Big \vert  \sum \limits_{j \in \phi _k  }  x_{j,m}[k] - x_{j,p}[k] \Big \vert^{2}}{4 N_0} }\right )^{-1}}.
\end{equation} 

\textit{Remark 2:} For  $N_{0}\to 0$ and $N_{0} \to \infty$,  $\mathcal  I_{LB}$  approaches  to $-K \left( {1} / {\ln2}- 1 \right)$ and $J\log M  -K \left( {1} / {\ln2}- 1  \right)$, respectively. This indicates that at   low and high signal-to-noise ratio (SNR) regions, there exists a constant gap   $-K \left( {1} / {\ln2}- 1 \right)$ between   $ \mathcal I_{LB}$ and   $\mathcal  I_{AMI} ^{\boldsymbol {\mathcal X}}$. The lower bound with a constant shift can well  approximate   the AMI, particularly in the low and high  SNR regions. In   medium SNR region, the gap between  the lower bound and AMI is intractable. In fact, maximizing the lower bound is still an efficient approach to improve the AMI in the medium SNR region \cite{LowYang,JiangLow}.



 
 \subsection{Error Performance  Analysis of the Proposed SSD-SCMA}
 \label{PEPsection}
 
 Assume  that the  erroneously decoded codeword is $\hat{\mathbf{w}}$ when  $ \mathbf{w}$ is transmitted, where  $ \hat{\mathbf{w}} \neq \mathbf{w}$.  Furthermore, let us define the element-wise distance $\tau _{{{\mathbf w}} \rightarrow {\hat{\mathbf{w}}}}(k) =   \vert w_k - \hat w_k \vert^2 $ \cite{liu2021sparse}.
Then, the  pairwise-error probability conditioned on  the  channel fading vector  for a maximum-likelihood receiver  is given as \cite{liu2021sparse}
\begin{equation}
\label{PEP1}
 \small
\text{Pr} \{\mathbf{w} \to \mathbf{\hat{w}} | \mathbf{H} \}  =    Q\left(\sqrt {\frac{ \sum\nolimits_{k = 1}^{2K}  h_k^2 \tau _{{{\mathbf w}} \rightarrow {\hat{\mathbf{w}}}}(k)}{2N_0}} \right),
\end{equation}
where  $ Q(t)=\frac{1}{\pi}\int _{0}^{+\infty } \frac{e^{ -\frac{x^2 }{2}(t^2+1)}  }{t^2+1}dt 
$ is the Gaussian $ Q $-function \cite{BoutrosGood}. 
By letting  $\gamma_{{{\mathbf w}} \rightarrow {\hat{\mathbf{w}}}}(k) = \tau _{{{\mathbf w}} \rightarrow {\hat{\mathbf{w}}}}(k) /{4N_0}$ 
and   with the expectation over the channel vector, one has 
\begin{equation}
\label{PEP_int}
\small
\begin{aligned}
      &  \text{Pr} \{\mathbf{w} \to \mathbf{\hat{w}}  \}  \\
          &= \frac{1}{\pi} \int _{0}^{+\infty } \frac{ \mathbb{E}_{\mathbf H}  \left \{ \exp \Bigg\{  - (t^2+1)\sum\limits_{k = 1}^{2K}  h_k^2 \gamma _{{{\mathbf w}} \rightarrow {\hat{\mathbf{w}}}}(k) \Bigg\} \right \}  } {t^2+1}dt \\
        &  = \frac{1}{\pi}\int _{0}^{+\infty } \frac{1}{t^2+1}    \prod _{k=1}^{2K} \frac {1}{1+ { \gamma _{{{\mathbf w}} \rightarrow {\hat{\mathbf{w}}}}(k) }    (t^2+1)}  dt. 
\end{aligned}
\end{equation}

 To proceed,  define the set $\eta_{{{\mathbf w}} \rightarrow {\hat{\mathbf{w}}}} \triangleq\left \{{k: \tau _{{{\mathbf w}} \rightarrow {\hat{\mathbf{w}}}}(k)\neq 0, 1\leq k \leq 2K }\right \}$, and let  $G_{{{\mathbf w}} \rightarrow {\hat{\mathbf{w}}}}$  be the  cardinality of   $\eta_{{{\mathbf w}} \rightarrow {\hat{\mathbf{w}}}}$. 
We further define $\delta_k \triangleq {\frac{\gamma _{{{\mathbf w}} \rightarrow {\hat{\mathbf{w}}}}(k)}{1+\gamma _{{{\mathbf w}} \rightarrow {\hat{\mathbf{w}}}}(k)}}$ and $\delta_{\text {min}} = \text {min} \left\{ \delta_k : k \in \eta_{{{\mathbf w}} \rightarrow {\hat{\mathbf{w}}}} \right\}  $.   Thus, the PEP can be written as \cite{PerformanceTaricco}
\begin{equation}
\label{PEP_final}
\small
  \text{Pr} \{\mathbf{w} \to \mathbf{\hat{w}}  \}     \leq B(G_{{{\mathbf w}} \rightarrow {\hat{\mathbf{w}}}}, \delta _{\text {min}}  ) \prod _{k=1}^{2K} \frac {1}{1+ { \gamma _{{{\mathbf w}} \rightarrow {\hat{\mathbf{w}}}}(k) } ^{2}  },
\end{equation}
where 
\begin{equation}
\small
\begin{aligned}
& B(G_{{{\mathbf w}} \rightarrow {\hat{\mathbf{w}}}}, \delta _{\text {min}}  )  \\
& =  \frac{1}{4^{G_{{{\mathbf w}} \rightarrow {\hat{\mathbf{w}}}}} }  \sum\limits_{l=0}^{G_{{{\mathbf w}} \rightarrow {\hat{\mathbf{w}}}}-1} 
  \begin{pmatrix}
 G_{{{\mathbf w}} \rightarrow {\hat{\mathbf{w}}}}-1+l \\
G_{{{\mathbf w}} \rightarrow {\hat{\mathbf{w}}}}
\end{pmatrix} 
 \left( \frac{2}{1+\delta _{\text {min}}}\right)^{G_{{{\mathbf w}} \rightarrow {\hat{\mathbf{w}}}}-l}.
\end{aligned}
\end{equation}

 Note that under the asymptotic condition $N_{\text{0}} \rightarrow 0$, (\ref{PEP_final}) is tighter than the Chernoff bound\footnote{The Chernoff bound can be obtained by applying   $B(G_{{{\mathbf w}} \rightarrow {\hat{\mathbf{w}}}}, \delta _{\text {min}}  ) = \frac{1}{2}$.
} by the factor $B(G_{{{\mathbf w}} \rightarrow {\hat{\mathbf{w}}}}, \delta _{\text {min}} )$.

\textit{Diversity order:}
At sufficiently  large SNR, since $\gamma_{{{\mathbf w}} \rightarrow {\hat{\mathbf{w}}}}(k)\rightarrow  \infty$, thus $\delta _{\text {min}} \approx   1 $  and it follows that 
\begin{equation}
\small
B(G_{{{\mathbf w}} \rightarrow {\hat{\mathbf{w}}}}, \delta _{\text {min}}  )  \leq \frac{1}{4^{G_{{{\mathbf w}} \rightarrow {\hat{\mathbf{w}}}}} } 
\begin{pmatrix} 2G_{{{\mathbf w}} \rightarrow {\hat{\mathbf{w}}}} -1 \\ G_{{{\mathbf w}} \rightarrow {\hat{\mathbf{w}}}} \end{pmatrix} .
\end{equation}
Hence, one can  approximate the right-hand side of (\ref{PEP_final}) as
 \begin{equation}
 \label{pep_final}
 \small
 \text{Pr} \{\mathbf{w} \to \mathbf{\hat{w}}  \} \leq
G_{c} \left(\mathbf {w}\rightarrow \hat{\mathbf {w}} \right)
  N_0^{{G_{d} (\mathbf {w}\rightarrow \hat {\mathbf {w}})}}, 
 \end{equation}
  where   
  \begin{equation}
\label{Gc_1}
\small
 G_{c} (\mathbf {w}\rightarrow \hat {\mathbf {w}})  =
\begin{pmatrix} 2G_{{{\mathbf w}} \rightarrow {\hat{\mathbf{w}}}} -1 \\ G_{{{\mathbf w}} \rightarrow {\hat{\mathbf{w}}}} \end{pmatrix} 
  \prod \limits _{k \in \eta_{{{\mathbf w}} \rightarrow {\hat{\mathbf{w}}}}}      {\left| \tau _{{{\mathbf w}} \rightarrow {\hat{\mathbf{w}}}}(k) \right|} ^{-2}.
  \end{equation}
 
 From (\ref{pep_final}) and (\ref{Gc_1}), define the diversity order (DO) of the SSD-SCMA system as       $G_{d}\triangleq \underset{\mathbf {w}\neq \hat {\mathbf {w}}}{\min } G_{d} (\mathbf {w}\rightarrow \hat {\mathbf {w}})$ and the coding gain as  $G_{c} \triangleq  \underset{\mathbf {w}\neq \hat {\mathbf {w}},  G_{d} (\mathbf {w}\rightarrow \hat {\mathbf {w}})=G_d }{\min}   G_{c} (\mathbf {w}\rightarrow \hat {\mathbf {w}})$ \cite{xin2003space}. 
The union bound of  average bit error rate (ABER) for SSD-SCMA system is given as
  \begin{equation}
  \label{ABER_}
  \small
  \begin{aligned} P_{\text {b}} \leq \frac {1}{M^{J} \cdot J\log _{2}(M)}  \sum _{\mathbf {w}}\sum _{\hat {\mathbf {w}}\neq \mathbf {w}}{n_{\text {E}}(\mathbf {w},\hat {\mathbf {w}})}\cdot \text{Pr} \{\mathbf{w} \to \mathbf{\hat{w}}\}. \\
  \end{aligned}
    \end{equation}
where  $n_e\left(\mathbf {w},\hat {\mathbf {w}}\right)$ is the Hamming distance between $\mathbf {w} $ and $\hat {\mathbf {w}}$.
Based on  (\ref{pep_final}) and (\ref{ABER_}), we introduce the following lemma:
 
 \textbf{Lemma 3:} The proposed SSD-SCMA enjoys maximum  DO of $G_d = 2N$, whereas a C-SCMA system only attains   DO of at most   $G_d = N$.    
 
 \textit{Proof:}
 In general, the maximum PEP among all codeword pairs  dominates the   ABER at high SNR values which corresponds  to the nearest neighbours among the multi-user codeword pairs. In other words, $ G_{d} (\mathbf {w}\rightarrow \hat {\mathbf {w}}) = G_{d}$  holds for the single error event that all the codewords are detected correctly except for  a codeword of only one user.  Hence, the proposed SSD-SCMA system can achieve maximum DO  of $2N$.  Specifically, the diversity comes from the following two aspects: whilst the non-zero elements of the sparse codebooks provide a diversity of $N$ (same as the traditional SCMA), the delay in the I/Q component attributes to a diversity increase of two times.

 
 
\textit{Remark 3:}  The proposed SSD-SCMA with a DO of $2N$ generally enjoys improved error performance with steeper ABER slope against SNR,   compared to C-SCMA system.   Observed from (\ref{pep_final}) and (\ref{ABER_}), it is ideal to maximize the DO and   coding gain of  a codebook  to improve ABER. In addition, the PEP  is also    dependent on    the element-wise distance $\tau _{{{\mathbf w}} \rightarrow {\hat{\mathbf{w}}}}(k)$,  hence it is also desirable to improve    $\tau _{{{\mathbf w}} \rightarrow {\hat{\mathbf{w}}}}(k)$ for improved  ABER.

 \section{Multidimensional codebook design:  design metrics}
 In this section, we formulate the  codebook design metrics according to the AMI  and PEP analysis.   The main idea is to maximize  the lower bound of AMI or  minimize  the upper bound of  PEP.

  \subsection{  AMI-CB Design by Maximizing the  Lower Bound of AMI}
Our optimization goal is to maximize the AMI lower bound derived in Lemma 2. Thus, we formulate the codebook design of  the SSD-SCMA system  with structure of ${\cal S}({\cal V},{\cal G}; J, M, N, K), \, \ {\cal V}: =[{\bf V}_{j}]_{j=1}^{J}$ and ${\cal G}:=[g_{j}]_{j=1}^{J}$ in downlink Rayleigh fading channels as follows:  
 \begin{subequations}
 \small
 \label{opt1}
 \begin{align}
   \mathcal {P}_{1} : {\cal V}^{\ast},{\cal G}^{\ast}  = & \quad \arg\max\limits_{{\cal V},{\cal G}}   \quad  \mathcal  I_{LB}^{\boldsymbol {\mathcal X}}  \tag{\ref{opt1}} \\
         \text {Subject to}   \quad   &   \quad 
          \sum\limits_{j=1}^{J}\text{Tr}\left( \boldsymbol{\mathcal X_j^H }\boldsymbol{\mathcal X_j }\right) = MJ.
     \end{align}
 \end{subequations}


\textit{1) MC figure of merit:} In general,   users'  sparse codebooks $\boldsymbol {\mathcal {X}}_j, j=1,2,\ldots,J$ are generated from a common MC, denoted as  $\boldsymbol {\mathcal C}_{MC} \in \mathbb C^{N \times M}$, thus the design of MC is crucial. The MC based MI (MC-MI)  determines the maximum    information rate that can be reliably transmitted    for a single user. Naturally,  we target maximizing the AMI lower bound  of the MC, which can be expressed as
\begin{equation}
\small
\begin{aligned} 
& \mathcal I_{{LB}}^{ \boldsymbol {\mathcal C}_{MC}} =   \mathrm {log}_{2}(M)  - \frac{N}{\ln2}    \\
       &  \quad \quad - \frac{1}{{{M}}} \log \left(  \sum\limits_{m=1}^{{{M}}}{ \sum\limits_{p=1}^{{{M} }}{\prod _{n=1}^{N} \left ({ 1 + \frac {\vert  {c}_{m,n}- {c}_{p,n}\vert ^{2}}{4N_{0}} }\right )^{-1}}  } \right),
\end{aligned}
\end{equation}
where $ {c}_{m,n}$ is the $m$th codeword at the  $n$th entry of $\boldsymbol {\mathcal C}_{MC}$.
Hence, in Rayleigh fading channels, we choose the following   cost function  for MC design: 
 \begin{equation}
 \small
 \label{MI_MC}
 \mathfrak {G}_{j} =   \sum\limits_{m=1}^{{{M}}}{ \sum\limits_{p=1}^{{{M} }}{\prod _{n=1}^{N} \left ({ 1 + \frac {\parallel  {c}_{m,n}- {c}_{p,n}\parallel ^{2}}{4N_{0}} }\right )^{-1}}}.
 \end{equation}
 

\textit{2) Labeling figure of merit:} 
Labeling is crucial for  BICM-IDD  system.  The effect of mapping on the performance of SSD-SCMA in BICM-IDD systems can be  characterized by the simplified criterion \cite{YangBitlabeling,bao2017bit}:
\begin{equation}
\small
\label{MI_label}
\begin{aligned}
\Upsilon _{\mathrm{ Ray}} = & \frac {1}{m M} \sum _{i=1}^{m} \sum _{b=0}^{1} \sum _{\mathbf {x} \in \mathcal {X}^{b}_{i}} \prod _{k=1}^{K} \Bigg \{ \frac {1}{ 1 + \frac {1}{4N_{0}} \vert \mathfrak{Re} \left( x_{k}-\hat x_{k}\right)  \vert^{2} }  \\ 
& \quad \quad \quad \quad \quad   \quad \times \frac {1}{ 1 + \frac {1}{4N_{0}} \vert \mathfrak{Im} \left( x_{k}-\hat x_{k}\right)  \vert^{2} } \Bigg \}, 
\end{aligned}
\end{equation}
where $ \mathcal  {X}^{b}_{i}$ denotes the set of codewords that with bit $b$ at $i$th position. Note that (\ref{MI_label}) is employed as the cost function to design the labeling for the AMI-CBs.

  \subsection{   P-CB  Design by Minimizing the Upper  Bound of  PEP}
   
  According to the \textit{Remark 3}  in Subsection \ref{PEPsection}, the ABER is dominated by the DO and the coding gain, i.e., $G_{c}$. The DO of user  $j$   is given by   
 \begin{equation} 
 \small
 \begin{aligned}
\text{DO}({\boldsymbol {\mathcal X}_{j}}) = &\sum \limits _{k=1}^{K}\text {Ind}\left (\vert \mathfrak{Re} \left(x_{n,i} -x_{n,l} \right)\vert \right) \\
& \quad \quad \quad + 
  \sum \limits _{k=1}^{K}\text {Ind}\left (\vert \mathfrak{Im} \left(x_{n,i} -x_{n,l} \right)\vert \right), 
 \end{aligned}
\end{equation}
  where $\text {Ind}(x)$ takes the value of one if $x$ is nonzero and zero otherwise. We now formulate the design metric by maximizing the coding gain. For any arbitrary $  {\mathbf {w}}$ and $ \hat {\mathbf {w}}$ that  $  \underset{\mathbf {w}\neq \hat {\mathbf {w}}}{\min } G_{d} (\mathbf {w}\rightarrow \hat {\mathbf {w}})= G_{d}$, the term $\prod \limits _{k \in \eta_{{{\mathbf w}} \rightarrow {\hat{\mathbf{w}}}}}      {\left| \tau _{{{\mathbf w}} \rightarrow {\hat{\mathbf{w}}}}(k) \right|} ^{-2}$ in (\ref{Gc_1}) equals to the product distance of a single user \cite{liu2021sparse}. To proceed, let us define the  modified MPD (MMPD) of the  $j$th user  as
 \begin{equation} 
 \small
 \begin{aligned}
d^{\boldsymbol {\mathcal X}_{j}}_{\mathrm {MMPD}} &= \underset {i\neq l  , 1<i,l<M}{\min }  \quad   d^{\boldsymbol {\mathcal X}_{j}}_{\mathrm {P},i,l}\\
& = \underset {i\neq l, 1<i,l<M}{\min } \prod _{n \in \rho_1(\mathbf {x}_i, {\mathbf {x}}_l)}  \Big \{ \vert \mathfrak{Re} \left(x_{n,i} -x_{n,l} \right)\vert ^{2} \\
&  \quad  \quad \quad \quad  \times  \prod _{n \in \rho_2(\mathbf {x}_i, {\mathbf {x}}_l)} \vert \mathfrak{Im} \left(x_{n,i} -x_{n,l} \right)\vert ^{2}\Big \},
 \end{aligned}
\end{equation}
where $\rho_1(\mathbf {x}_i, {\mathbf {x}}_l)$ and $\rho_2(\mathbf {x}_i, {\mathbf {x}}_l)$  denote the   sets of indices in which $ \mathfrak{Re} ( {{x}}_{n,i} )\neq  \mathfrak{Re} ({{x}}_{n,j})$ and  $ \mathfrak{Im} ( {{x}}_{n,i} )\neq  \mathfrak{Im} ({{x}}_{n,j})$, respectively.      
Note that  the product distance of one user may be different from that of another    due to different user  constellation operators. Hence, it is interesting to minimize  the MMPD of all users, which is  expressed as
 \begin{equation} 
 \small
 \label{MMPD}
d^{\boldsymbol {\mathcal X} }_{\mathrm {MMPD}} = \underset{j=1,2,\ldots,J}{\text{min}}  \; \; \; d^{\boldsymbol {\mathcal X}_{j}}_{\mathrm {MMPD}}.  
\end{equation}
Obviously,  improving the $G_{c}$   of a codebook is equivalent to maximizing  the $d^{\boldsymbol {\mathcal X} }_{\mathrm {MMPD}}$. In addition, we further define the minimum  element-wise distance   $\tau_{\mathrm {min}}^{\boldsymbol {\mathcal X}} $ as 
 \begin{equation}
 \label{tmin}
 \small
 \begin{aligned}
\tau_{\mathrm {min}}^{\boldsymbol {\mathcal X}} = &  \min \big\{  \tau^2_{{{\mathbf w}} \rightarrow {\hat{\mathbf{w}}}}(k) \\ & \vert 
   \forall {{\mathbf{w}}_{n}},{{\mathbf{w}}_{m}}\in \Phi,   {{\mathbf{w}}_{n}} \neq {{\mathbf{w}}_{m}}, 1 \leq k \leq 2K \big\}.
 \end{aligned}
  \end{equation}

Based on the \textit{Remark 3}   in Subsection    \ref{PEPsection}, it is desirable to design codebooks  to achieve full DO, large MMPD and large $\tau_{\mathrm {min}}^{\boldsymbol {\mathcal X}}$. Hence, the codebook design problem of SSD-SCMA is formulated as 
 \begin{subequations}
 \small
 \label{opt2}
 \begin{align}
    \mathcal {P}_{2-1} :  {\cal V}^{\ast},{\cal G}^{\ast} = & \arg\max\limits_{{\cal V},{\cal G}}  \left\{d^{\boldsymbol {\mathcal X}  }_{\mathrm {MMPD}} ,   \tau_{\mathrm {min}}^{\boldsymbol {\mathcal X}  }  \right\}  \tag{\ref{opt2}}\\
        \text {Subject to} \quad   &   
          \text{DO}({\boldsymbol {\mathcal X}_{j}}) = 2N, j= 1,2,\ldots,J, \label{opt2:a} \\
        &     \sum\limits_{j=1}^{J} \text{Tr}\left( \boldsymbol{\mathcal X_j^H }\boldsymbol{\mathcal X_j }\right) = MJ.   \label{opt2:b}
     \end{align}
 \end{subequations}

\textit{1) MC figure of merit:}

Similar to the case of  MI aided MC design, the MC constrained PEP is  employed to guide the   design of  MC. Specifically,  the MC design is formulated as  
\begin{equation} 
\small
\label{PEP_MC}
 \begin{aligned}
\boldsymbol {\mathcal C}_{MC}^{\ast}  = \;& \text{max} \; \;   \underbrace{    \underset {i\neq l}{\min } \prod _{n \in \rho(\mathbf {c}_i, {\mathbf {c}}_l)} \vert c_{n,i} -c_{n,l} \vert ^{2}}_{d^{\boldsymbol {\mathcal C}_{MC}}_{\mathrm {MPD}}}, \\
&  \text {Subject to}\; \;   \begin{array}{l}
        \text{Tr} \left( \boldsymbol {\mathcal C}_{MC}^H \boldsymbol {\mathcal C}_{MC}  \right)= NM,  \\
       \vert \rho (\mathbf {c}_i, {\mathbf {c}}_j) \vert = N,
\end{array} 
\end{aligned}
\end{equation}
 where $\rho (\mathbf {c}_i, {\mathbf {c}}_j)$ denotes the set of  indices in which $ {c}_{n,i} \neq {{c}}_{n,j}$, and $d^{\boldsymbol {\mathcal C}_{MC}}_{\mathrm {MPD}}$ is the MPD of $\boldsymbol {\mathcal C}_{MC}$.   The rationale for employing $d^{\boldsymbol {\mathcal C}_{MC}}_{\mathrm {MPD}}$ is that  a  MC with large  $d^{\boldsymbol {\mathcal C}_{MC}}_{\mathrm {MPD}}$   can also enlarge  $d^{\boldsymbol {\mathcal X}_{j}}_{\mathrm {MMPD}} $ after applying user-specific operation, such as phase rotation. In addition,  it is well known that improving $d^{\boldsymbol {\mathcal C}_{MC}}_{\mathrm {MPD}}$ can also improve the ABER of a C-SCMA system.

\textit{2) Labeling figure of merit:}  The  labeling rules should be  designed to minimize the MC constrained PEP. Under Rayleigh fading channels, the labeling  metric is given as \cite{chen2020design}
\begin{equation}
\small
\label{label_pep}
\Pi_{\text{R}} (\xi _{l})  \equiv \sum _{i=1}^{M-1}   \sum _{l=i+1}^{M} N_{i,l}(\xi _{l}) \frac {1}{ d^{\boldsymbol {\mathcal X}_{j}}_{\mathrm {P},i,l}},
\end{equation}
where $N_{i,l}(\xi _{j})$ denotes  the number of different labelling bits between $\mathbf {x}_{j,i} $ and  $\mathbf {x}_{j,l} $ based on the considered labelling rule $\xi _{j}$. 

 \subsection{Asymptotic Relationship  Between the P-CB and AMI-CB }
 
 \label{SymboticRela}
 
First, let us look at    the AMI lower bound, i.e., $ \mathcal  I_{LB}^{\boldsymbol {\mathcal X}}$ in   (\ref{AMI_LR}),  whose value   depends on  $      \sum\nolimits_{m=1}^{{{M}^{J}}} \sum\nolimits_{p=1}^{{{M}^{J}}} $ ${\prod _{k=1}^{K}\lambda _{k,m,p}}$.  For sufficiently    high SNR, we have 
\begin{equation}
\small
     {  \gamma _{k,m,p} \approx  \prod _{l \in \{ \mathrm I, \mathrm Q\}}  \Bigg ({  \frac {  \Big \vert  \sum \limits_{j \in \phi _k }  x_{j,m,l}[k] - x_{j,p,l}[k] \Big \vert^{2}}{4N_{0}} }\Bigg )^{-1}}.
\end{equation}
The term  $      \sum\nolimits_{m=1}^{{{M}^{J}}}{ \sum\nolimits_{p=1}^{{{M}^{J}}}{\prod _{k=1}^{K}     \gamma _{k,m,p}}}$ is dominated by the most significant term. Thus,   for $N_0 \rightarrow 0$, we have the following lower bound:
 \begin{equation}
 \small
 \label{MIPEP}
 \begin{aligned}
\sum\limits_{m=1}^{{{M}^{J}}}{ \sum\limits_{p=1}^{{{M}^{J}}}{\prod _{k=1}^{K}     \gamma _{k,m,p}}} & \geq \underset{1\leq m,p \leq M^{J}} { \max} \quad  \frac{1}{M^J} \prod _{k=1}^{K}  \gamma _{k,m,p}.
 \end{aligned}
 \end{equation}
 
 It is noted  that the right-hand side of (\ref{MIPEP})   essentially corresponds to the MMPD of the codebook, i.e., the $d^{\boldsymbol {\mathcal X} }_{\mathrm {MMPD}}$ in (\ref{MMPD}).
Hence,   maximizing $\mathcal  I_{LB}^{\boldsymbol {\mathcal X}}$ of a codebook is equivalent to maximizing      the $d^{\boldsymbol {\mathcal X} }_{\mathrm {MMPD}}$ of a codebook at a sufficiently high SNR value. 

\textit{Remark 4 :} Similar  results can be obtained for   C-SCMA system. Namely, a codebook with larger MPD will result  in a  higher $\mathcal  I_{LB}^{\boldsymbol {\mathcal X}}$ with  $\gamma _{k,m,p}$ given in (\ref{CSCMAMI}) at a sufficiently high SNR values. 

 \section{Multidimensional codebook design: implementation issues}
 
This section presents the detailed sparse codebook design for SSD-SCMA based on the proposed design metrics in Section IV.   Specifically,
Subsection   \ref{SecCSS}  first introduces a      constellation superposition scheme,  which generates multiple sparse codebooks based on the MC.  Then, we present the  solutions for $\mathcal {P}_{1-1}$ and $\mathcal {P}_{1-2}$, i.e., (\ref{opt1}) and (\ref{opt2}),   in Subsections \ref{MI_CB} and  \ref{PEP_CB}, respectively.

 \subsection{Constellation Superposition Scheme}
 \label{SecCSS}
As mentioned, multiple sparse codebooks are generated from a common MC  ($\boldsymbol {\mathcal C}_{MC} \in \mathbb C^{N \times M}$) by  user-specific
operations.  The detail design of  $\boldsymbol {\mathcal C}_{MC}$ will be discussed later.  Once $\boldsymbol {\mathcal C}_{MC}$ is determined,  phase rotations are applied to design  the multi-dimensional constellations  for different users.  Therefore, the $j$th user's codebook with non-zero elements is generated as  $ \boldsymbol {\mathcal C}_{j} =   \mathbf{R}_j  \boldsymbol {\mathcal C}_{MC}$, where $\mathbf{R}_j$  denotes the diagonal     phase rotation matrix of the $j$th user.  For example, for an $\boldsymbol {\mathcal C}_{MC}$ with  $N=2$,  we have $\mathbf{R}_j=\left[ \begin{matrix}
   e^{j \theta _1}   &  0 \\
     0   &     e^{j \theta _2} 
\end{matrix} \right]$. Based on the mapping matrix $\mathbf{V}_j$, the $j$th user's codebook now can be  generated by $\boldsymbol{\mathcal X}_{j} = \mathbf {V}_{j} \mathbf{R}_j \boldsymbol {\mathcal C}_{MC}$.  $\mathbf {V}_{j}$ can be constructed based on the $j$-th column of $\mathbf{F}$. Specifically, according to the positions of the ‘$0$’ elements of  ${{\mathbf{f}}_{j}}$, we insert the all-zero row vectors into the identity matrix ${{\mathbf{I}}_{N}}$. 
 For example, for the  $\mathbf{F}$    in Fig. \ref{factor}, we have
\begin{equation} 
\small
{{\mathbf{V}}_{1}}=\left[ \begin{matrix}
  0 & 0  \\
  1 & 0  \\
  0 & 0  \\
  0 & 1  \\
\end{matrix} \right],{{\mathbf{V}}_{2}}=\left[ \begin{matrix}
  1 & 0  \\
  0 & 0  \\
  0 & 1  \\
  0 & 0  \\
\end{matrix} \right].
 \end{equation}

It is noted that  the phase rotation matrix   $\mathbf{R}_j$  and the mapping  matrix $\mathbf {V}_{j}$ can be combined together by a column vector, i.e., $\mathbf{s}_{N \times J}^j =\mathbf {V}_{j} \mathbf{R}_j \mathbf{I}_K$, where $\mathbf{I}_K$ denotes all-one vector of length $K$.   Hence, the codebooks for the $J$ users  can be represented by the signature matrix  $\mathbf{S}_{N \times J}= \left[ \mathbf{s}_{N \times J}^1, \mathbf{s}_{N \times J}^2, \dots, \mathbf{s}_{N \times J}^J \right]$. In this paper, we consider the following     signature matrix: 

  \begin{equation} 
 \label{signature_46}
 \small
 {{\mathbf{S}}_{4\times 6}}=\left[ \begin{matrix}
   0  & e^{j\theta_3} & e^{j \theta_1} & 0 &  e^{j\theta_2} & 0  \\
   e^{j\theta_2} & 0 & e^{j\theta_3} & 0 & 0 & e^{j\theta_1}\\
   0 & e^{j\theta_2}& 0 & e^{j\theta_1} & 0 & e^{j\theta_3} \\
   e^{j\theta_1} & 0 & 0 & e^{j\theta_2} & e^{j\theta_3} & 0  \\
\end{matrix} \right].
  \end{equation}
 
  \begin{figure}[]
  \centering
  \includegraphics[width=3.5in]{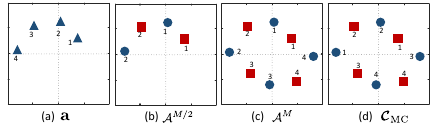} 
  \caption{An MC example  based on GAM, where $\xi =1$, $\psi = -0.74$, $M=4$, $N=2$. The elements in (a) are the original GAM points, i.e., $\mathbf a$. Subfigures (b) to (d) show the mapping process from $\mathbf a$ to $\boldsymbol {\mathcal C}_{MC}$.  Square  and circular marks are the elements of the first and second dimensions, respectively.  }
\label{GAM_MC}
\end{figure}

\subsection{Design of AMI-CB} 
\label{MI_CB}

 \begin{algorithm}[t]  
\caption{Design $\boldsymbol {\mathcal C}_{MC}$ based on GAM.}
\label{GAM_al}
\begin{algorithmic}[1]
\REQUIRE{  $\xi, \psi$, $N$, $M$ and $N_0$}\\
\STATE \textbf{Step 1. Generate the basic constellation}  
\STATE  $\mathbf a$ $\gets$   Generate $a_l = c_{\text {norm}}\sqrt{l+\xi} e^{i2\pi (\varphi + \psi)L}$, $l=1,2,\ldots, \frac{MN}{2}$  
  \FOR{$l=1:\frac{MN}{2}$ } 
  \STATE $n = \text{mod}(l,N)$
\STATE   $\mathcal A^{\frac{M}{2}}$ $\gets$  map the $l$th element of $\mathbf a$ to the $n$th dimension of  $\mathcal A^{\frac{M}{2}}$
\ENDFOR 
\STATE    $\mathcal A^{ {M} }$ $\gets$  $\left( \mathcal A^{\frac{M}{2}}, -\mathcal A^{\frac{M}{2}} \right) $  \\
\STATE \textbf{Step 2. Dimension permutation} \\
\STATE $\boldsymbol {\mathcal C}_{MC} $ $\gets$   $ \left[ {\pi }_{1} \left( \mathcal A_{1}^{{M}} \right),{\pi }_{2} \left( \mathcal A _{2}^{{M}} \right),\ldots,{\pi }_{N} \left( \mathcal A _{N}^{{M}} \right)\right]^{\mathcal T}$ based on the criteria given in (\ref{MI_MC})\\
\STATE \textbf{Step 3. Dimension switching} \\
\STATE $\boldsymbol {\mathcal C}_{MC}^{'}= \left[\mathbf c_N^{\mathcal T}, \mathbf c_{N-1}^{\mathcal T}, \ldots, \mathbf c_1^{\mathcal T} \right]^{\mathcal T}$
\end{algorithmic}
\end{algorithm}

 \textit{1) Proposed Design of  $\boldsymbol {\mathcal C}_{MC}$:}  GAM is a novel, shape-versatile and circular symmetric   modulation scheme which can offer enhanced MI  performance over pulse-amplitude modulation  and square quadrature amplitude modulation (QAM) design.  
 This motivates us to employ GAM as the basic MC  to design the AMI-CBs. In an  $N_p$-point disc-shaped GAM,   the $n^{th}$ constellation point  can be  generated according to $a_n = r_ne^{i2\pi \varphi n}$, where $r_n = c_{\text {norm}}\sqrt{n}$, $c_{\text {norm}} = \scriptstyle\sqrt{\frac{2P}{N_p+1}}$,  $P$ is the  power constraint and $\varphi = \frac{1-\sqrt{5}}{2}$ is the golden angle in rads.  The ($\xi, \psi$)-GAM is defined as  $a_n = c_{\text {norm}}\sqrt{n+\xi} e^{i2\pi (\varphi + \psi)n}$ \cite{mheich2018design}, where $\xi$ and  $\psi$ are the two parameters to be optimized in the MC. Here, we propose an   enhanced  scheme to construct the MC  based on GAM, termed as E-GAM. The design scheme mainly includes the following three steps and  the detailed process is given in \textbf{Algorithm \ref{GAM_al}}.
 
\textit{ Step 1}. Generate the $N$-dimensional constellation based on GAM, denoted as $ \mathcal A^{ {M} }= [ \mathbf{a}_1^{\mathcal T}, \mathbf{a}_2^{\mathcal T}, $ $ \ldots,  \mathbf{a}_N^{\mathcal T}]^{\mathcal T} \in \mathbb C^{N \times M} $.
 
 \textit{Step 2}. Perform permutation to obtain higher gain based on the criteria given in (\ref{MI_MC}).    For $n=1, \ldots, N$, let $\boldsymbol {\pi }_{n}:\{1,\ldots,M\} \rightarrow \{1,\ldots,M\}$  denote the permutation mapping of   dimension $n$. Namely,   $ {\pi }_{n} \left( \mathbf{a}_n \right)$ is the permutation operation of $\mathcal A _{{M}}$ at $n$th dimension. Then the $N$-dimensional constellation can be obtained as 
\begin{equation}
\small
     \boldsymbol {\mathcal C}_{MC} = \left[ {\pi }_{1} \left( \mathbf{a}_1^{\mathcal T} \right),{\pi }_{2} \left( \mathbf{a}_2^{\mathcal T}\right),\ldots,{\pi }_{N} \left( \mathbf{a}_N^{\mathcal T} \right)\right]^{\mathcal T}.
\end{equation}

The construction of an  $N$-dimensional constellation is equivalent to  finding  the  $N$ permutations  ${\pi }_{n}, n=1,2, \ldots,N$,  which can be efficiently solved
by using the binary switching algorithm (BSA) \cite{zeger1990pseudo,bao2017bit}.


\textit{Step 3}. Dimension switching. Note that different  from  conventional codebook design  \cite{chen2020design}, where  a  basic one-dimensional  constellation is repeated at each   MC,  $\boldsymbol {\mathcal C}_{MC}$ is directly designed  based on GAM, which introduces inherent power difference among  the $N$ dimensions.  For the $n$th dimension, the total energy can be obtained as 
 \begin{equation}
 \small
     E_n = c_{\text {norm}}^2 M\left( \xi+n+  N\left( \frac{M}{2}- 1\right)   \right), 
 \end{equation}
 and the power difference   between two consecutive dimensions is  $  E_{n+1}-E_{n}=c_{\text {norm}}^2M, 1 \leq n\leq N-1$.  To further maintain the power difference of the sparse codebooks, dimension switching is introduced to   $\boldsymbol {\mathcal C}_{MC} $ for some users. Specifically, let $\mathbf c_n \in \mathbb C^{1 \times M}$ be the $n$th row of  $\boldsymbol {\mathcal C}_{MC} $,   the switched $\boldsymbol {\mathcal C}_{MC} $ can be written as 
  \begin{equation}
  \small
\boldsymbol {\mathcal C}_{MC}^{'}= \left[\mathbf c_N^{\mathcal T}, \mathbf c_{N-1}^{\mathcal T}, \ldots, \mathbf c_1^{\mathcal T} \right]^{\mathcal T}.
 \end{equation}

\textit {Remark 5}: The major differences between the proposed E-GAM and \cite{mheich2018design} are: (1) The proposed E-GAM can avoid complex mapping from the GAM points to MC  through mapping table; (2) A mathematical  design criteria is considered in the proposed E-GAM, which is missing in \cite{mheich2018design}; (3) A novel dimension switching scheme is further proposed to improve the power diversity.

 \textit{2) Generate users' sparse codebooks:} 
  Based on the $\boldsymbol {\mathcal C}_{MC}$ and $\boldsymbol {\mathcal C}_{MC}^{'}$, user $j$'s  sparse codebook is obtained as 
   \begin{equation}
   \small
   \label{cbGen}
   \boldsymbol{\mathcal X}_{j} =
\left\{ \begin{matrix} \mathbf {V}_{j} \mathbf{R}_j \boldsymbol {\mathcal C}_{MC}, \quad j  \; \text{is odd,} \\ 
\mathbf {V}_{j} \mathbf{R}_j \boldsymbol {\mathcal C}_{MC}^{'}, \quad j  \; \text{is even.} 
\end{matrix}\right.
 \end{equation}
 
 (\ref{cbGen}) introduces   row-based energy difference in   (\ref{signature_46}) by  $ E_1 \neq  E_2$, which is useful for improving  the distance  profile of the superimposed codewords.
 For example, according to (\ref{signature_46}),  users $2$, $3$, and $5$ superimpose over the first resource node with individual energy of $E_2$, $E_1$, and $E_1$, respectively. 
 
   The MC parameters, i.e., $\xi, \psi$,  and the rotation angle  $\theta_i, i=1,2,\ldots,  d_f$  should be optimized according to   (\ref{opt1}). Let $\boldsymbol \Theta  = \left [\theta_1,\theta_2, \ldots, \theta_{d_f}, \xi, \psi  \right]^{\mathcal T}$ denote all the parameters to be optimized.
Based on the proposed MC and constellation superposition scheme, the optimization problem $\mathcal {P}_{1-1}$ is reformulated as
 \begin{equation}
 \small
 \label{opt4}
 \begin{aligned}
  \mathcal{ P}_{1-2}:  {\boldsymbol {\mathcal X} } =\arg\max\limits_{ \boldsymbol \Theta }\; \; \;  v_{\text{Obj}}(\boldsymbol \Theta) = &  \quad  \mathcal  I_{LB}^{\boldsymbol {\mathcal X}} \\
      \text {Subject to}   \; \; \;
    v_{i}(\boldsymbol \Theta) =   & \left\{  \begin{array}{l}
    -\theta_i \leq 0 \\
      \theta_i-\pi \leq 0, i=1,2,\ldots, d_f   
       \end{array} \right. \\
      v_{d_f+1}(\boldsymbol \Theta) =   & \left\{  \begin{array}{l}
     - \frac{M}{2}-\xi \leq 0  \\
      \xi- \frac{M}{2}\leq 0  
       \end{array} \right. \\     
            v_{d_f+2}(\boldsymbol \Theta) =   & \left\{  \begin{array}{l}
    -\frac{\pi}{M}-\psi \leq 0  \\
      \psi- \frac{\pi}{M}  \leq 0 . 
       \end{array} \right. \\   
     \end{aligned}
 \end{equation}

  \begin{algorithm}[t]  
\caption{ AMI-CB design based  on the   AMI lower bound  $\mathcal I_{LB}^{\boldsymbol {\mathcal X}}$}
\label{CB_Mi}
\begin{algorithmic}[1]
\REQUIRE{$J$, $K$, $\mathbf V_j$, $M$,   $N_0$     }\\
\FOR{$i_1 = 1:I_1$}   
 \STATE  Randomly choose an initial value of  $\boldsymbol{\Theta}$ that satisfies the constraint.
\FOR{$i_2 = 1:I_2$}   
 \STATE  For the given    $\xi, \psi \in \boldsymbol{\Theta}$, generate    $\boldsymbol {\mathcal C}_{MC}$ based on \textbf{Algorithm \ref{GAM_al}}.\\
 \STATE  Perform dimension switching and generate ${\boldsymbol {\mathcal X}_j }$  according to (\ref{cbGen}). \\
 \STATE  Compute (\ref{HASS}) with the proposed sub-optimal estimation, and obtain the update direction $  \Delta   \boldsymbol{\Theta} ,  \Delta{ \mathbf u }$.\\
 \STATE   Line search method to determine the update factor $\varsigma$ by minimizing $\Vert \mathbf r_{\text{dul}}\Vert +  \Vert\mathbf r_{\text{cent}}\Vert$, and then update  $    \boldsymbol{\Theta} =  \Delta   \boldsymbol{\Theta}+ \varsigma \Delta   \boldsymbol{\Theta}$, $    \mathbf u =  \mathbf u+ \varsigma \Delta{ \mathbf u }$.\\
 \ENDFOR \\
  \STATE   Preserve the current results of $ \boldsymbol{\Theta} $ and  $v_{\text{Obj}}(\boldsymbol \Theta)$. \\
 \ENDFOR  \\
   \STATE Choose the best result $ \boldsymbol  \Theta^{*}$ that has the maximum value of $v_{\text{Obj}}(\boldsymbol \Theta^{*})$, and generate ${\boldsymbol {\mathcal X}_j }$ based on  $ \boldsymbol  \Theta^{*}$.  \\
   \STATE Perform the bit labeling for ${\boldsymbol {\mathcal X}_j }$ based on the criteria given in (\ref{MI_label}).
\end{algorithmic}
\end{algorithm}

Obviously, there are   $d_f +2$  parameters to be optimized in  (\ref{opt4}). Unfortunately,   $v_{\text{Obj}}(\boldsymbol \Theta)$ is a   non-convex function and the computation complexity of $ v_{\text{Obj}}(\boldsymbol \Theta)$ is  also high, especially for $8\leq M$.       To solve (\ref{opt4}), we propose a sample based primal-dual IPM with random initial values. We first transform    (\ref{opt4}) into a standard barrier problem with the  perturbed Karush–Kuhn–Tucker conditions  given by \cite{boyd2004convex}
 \begin{equation}
 \small
  \begin{aligned}
\nabla v_{\text{Obj}}(\boldsymbol \Theta) + \sum\nolimits_{p=1}^{d_f+2}u_p \nabla v_p(\boldsymbol \Theta) = 0,&\\
  u_p \nabla v_p(\boldsymbol \Theta) =-1/t,& \\
  v_{p}(\boldsymbol \Theta) \leq 0,  u_p \geq 0, &\\
   p =1,2,\ldots,d_f+2,&
     \end{aligned}
 \end{equation}
where $u$ is the Lagrangian factor and $t$ is the barrier factor.
In each iteration, the primal-dual update direction $  \Delta   \boldsymbol{\Theta} ,  \Delta{ \mathbf u }$ are obtained by iteration  solving the following equation
 \begin{equation}
 \label{HASS}
 \small
 \begin{aligned}
 \begin{bmatrix}
  \mathbf  H_{\text{es}} &  \nabla v(\boldsymbol \Theta) \\
 -\text{diag} \left( \mathbf u \right)\nabla v(\boldsymbol \Theta) &-\text{diag} \left( \mathbf v(\boldsymbol \Theta) \right)  
\end{bmatrix}
 \begin{bmatrix}
  \Delta   \boldsymbol{\Theta}  \\ \Delta{ \mathbf u }
\end{bmatrix} 
   = -\begin{bmatrix}
     \mathbf r_{\text{dul}}  \\  \mathbf r_{\text{cent}}
\end{bmatrix}  ,
 \end{aligned}
 \end{equation}
where $\mathbf u = [u_1, u_2, \ldots, u_{d_f+2}]^{\mathcal T}$, $ \mathbf H_{\text{es}}=\nabla^2 v_{\text{Obj}}(\boldsymbol \Theta) + \sum\nolimits_{p=1}^{d_f+2}u_p \nabla^2 v_p(\boldsymbol \Theta)  $ is the  Hessian matrix, $ \mathbf r_{\text{dul}}= \nabla v_{\text{Obj}}(\boldsymbol \Theta) + \sum\nolimits_{p=1}^{d_f+2}u_p \nabla v_p(\boldsymbol \Theta) $, and $\mathbf r_{\text{cent}}= -\text{diag} \left( \mathbf u \right)  v(\boldsymbol \Theta) - 1/t$.   In (\ref{HASS}), the estimation of $ \mathbf H_{\text{es}}$ is updated with the BFGS method  \cite{boyd2004convex}, i.e., 
 \begin{equation}
 \small
\mathbf H_{\text{es}}^{(t+1)} = \mathbf H_{\text{es}}^{(t)} + \frac{\mathbf q^{(t)} \mathbf q^{(t)T}}{\mathbf q^{(t)}   \Delta \boldsymbol{\Theta}^{(t)} }  - \frac{\mathbf H_{\text{es}}^{\mathcal T}  \Delta \boldsymbol{\Theta}^{(t)}  \Delta \boldsymbol{\Theta}^{(t)T}  \mathbf H_{\text{es}}^{(t)T}}{    \Delta \boldsymbol{\Theta}^{(t)T}   \mathbf H_{\text{es}}^{(t)} \Delta \boldsymbol{\Theta}^{(t)} },
 \end{equation}
where $\mathbf q^{(t)}  = \sum\nolimits_{p=1}^{d_f+2}u_p \left( \nabla v_p(\boldsymbol \Theta ^{(t+1)}) - \nabla v_p(\boldsymbol \Theta ^{(t)} )  \right)  + \nabla v_{\text{Obj}}(\boldsymbol \Theta ^{(t+1)}) - \nabla v_{\text{Obj}}(\boldsymbol \Theta ^{(t)})$.  The above iteration stops when the size of the current step is less than
the value of the step size tolerance or the constraints are 
satisfied to within the value of the constraint tolerance. Since the $\mathcal I_{LB}^{\boldsymbol {\mathcal X}}$ is a non-differentiable function, the difference method is employed to calculate the numerical gradient   $\nabla v_{\text{Obj}}(\boldsymbol \Theta)$. Due to the high computational complexity of calculating  $\nabla v_{\text{Obj}}(\boldsymbol \Theta)$ for  $M  \geq 8 $,  a sub-optimal Monte Carlo sample of    $\nabla v_{\text{Obj}}(\boldsymbol \Theta)$ for  $M  \geq 8 $ is used to estimate its true value.  In addition,  due to the non-convex of the objection function, the above process can be performed several times with different initial conditions to prevent it from converging  into local optimum.  The overall proposed  efficient codebook design  based on the AMI lower bound is summarized in Algorithm \ref{CB_Mi}.


 
\textit{3) Perform bit labeling:}  For each of the sparse codebook  $\boldsymbol{\mathcal X}_{j} $, we only construct its $M$ codewords for the codebook, the corresponding bit-to-symbol mapping, i.e., bit labeling, is further need to be assigned. For an $M$-ary codebook, there are  a total of $T=M!$ labeling methods.  The BSA can be employed again to find the labeling solution for  $M  \geq 8 $ with reasonable complexity, whereas the  labeling can be done by exhaustive search for the codebook set with small $M$, i.e., $M \leq 4$. 
 
 \subsection{Design of P-CB} 
 \label{PEP_CB}
  \begin{algorithm}[t] 
\caption{  P-CB design based  on  the   PEP upper bound in (\ref{pep_final})}
\label{CBPEP}
\begin{algorithmic}[1] 
\REQUIRE{$J$, $K$, $\mathbf V_j$, $M$,   $N_0$     }
\STATE    For    given $M$, choose a one-dimensional basic constellation $\mathbf a$ from  the constellation pool.\\
\STATE   Permute $\mathbf p$ to obtain   $\boldsymbol {\mathcal C}_{MC}$  according to the criteria given in (\ref{PEP_MC}). 
  
\STATE    From the perspective of the basic constellation $\mathbf p$, obtaining  the  optimal rotation angles  $\theta_{\text{opt}}^{1} \in (0, \frac{\pi}{2}]$ and $\theta_{\text{opt}}^{2} \in (-\frac{\pi}{2}, 0]$ that achieve the  largest DO and  MMPD, and let $\theta_{1} = \theta_{\text{opt}}^1$, $\theta_{2} = \theta_{\text{opt}}^2$.
\\
\STATE  From the perspective of the superimposed codewords, search  $\theta_{3} \in [-\pi, \pi]$  for the largest $\tau_{\mathrm {min}}^{\boldsymbol {\mathcal X}}$ in (\ref{delta1}).  
\STATE According to (\ref{signature_46}), generate sparse codebooks by $\boldsymbol{\mathcal X}_{j} = \mathbf {V}_{j} \mathbf{R}_j \boldsymbol {\mathcal C}_{MC}, j=1, 2, \ldots, J$.
\STATE   
Perform    bit labeling  for the optimized codebook $\boldsymbol{\mathcal X}_{j}$, and use the labeling metric given in (\ref{label_pep}). 
\end{algorithmic} 
\end{algorithm}
 \textit{1) Design of  $\boldsymbol {\mathcal C}_{MC}$ :} In contrast to the case for the AMI based MC design, a MC with large MPD, i.e, $d^{\boldsymbol {\mathcal C}_{MC}}_{\mathrm {MPD}}$, is required for the PEP based design. To this end, we first choose a one-dimensional basic constellation with large MED, denoted as $\mathbf p$, and then permute  $\mathbf p$  to obtain $\boldsymbol {\mathcal C}_{MC}$. Namely, the $N$-dimensional constellation can be obtained as 
\begin{equation}
     \boldsymbol {\mathcal C}_{MC} = \left[ {\pi }_{1} \left( \mathbf p\right),{\pi }_{2} \left( \mathbf p \right),\ldots,{\pi }_{N} \left(\mathbf p \right)\right]^{\mathcal T}. 
\end{equation}

 Similar to the  AMI based codebook design, the permutation is obtained by performing    BSA according to the criteria given in (\ref{PEP_MC}).   

 \textit{2) Generate   sparse codebooks:} Multiple  sparse codebooks are generated by $\boldsymbol{\mathcal X}_{j} = \mathbf {V}_{j} \mathbf{R}_j \boldsymbol {\mathcal C}_{MC}$   based on the signature matrix given in (\ref{signature_46}). Let $p_i$ denote the  $i$th signal point in $\mathbf p$. Based on (\ref{signature_46}), the superimposed constellation on a resource node can be obtained  by 
   \begin{equation}
 \small
  \begin{aligned}
 \boldsymbol{\mathcal S}_{\text{sum}}  =   \Big\{& p_i^{(1)} e^{j\theta_1} + p_i^{(2)} e^{j\theta_2} + \ldots  + p_i^{(d_f)}e^{j\theta_{d_f}} \\
 &  \quad \quad \quad \quad \vert \forall p_i^{(l)}\in  \mathbf {p}, l = 1,2,\ldots,d_f \Big\}.
      \end{aligned}
 \end{equation}
 
 Accordingly, the   $\tau_{\mathrm {min}}^{\boldsymbol {\mathcal X}} $ in  (\ref{tmin}) can be simplified to 
  \begin{equation} 
 \small
 \label{delta1}
 \begin{aligned}
\tau_{\mathrm {min}}^{\boldsymbol {\mathcal X}}  =   \underset{m \neq n} {\min}\big \{   \min \big \{   \vert \mathfrak{Re}( {s}_{m} -  {s}_{n}) \vert ^2,  \vert \mathfrak{Im}( {s}_{m} -  {s}_{n}) \vert ^2 \big\}\big\},
 \end{aligned}
  \end{equation}
 where ${s}_{m}$ is the $m$th point of $ \boldsymbol{\mathcal S}_{\text{sum}} $. According to (\ref{opt2}),  the rotation angles, i.e.,   $\boldsymbol{\theta} = \left [\theta_1. \theta_2, \ldots, \theta_{d_f}\right]$, should be  designed  to achieve full DO,   large  MMPD  $ \left(d^{\boldsymbol {\mathcal X}}_{\mathrm {P,min}}\right)$,  and  large  $\tau_{\mathrm {min}}^{\boldsymbol {\mathcal X}} $, i.e.,
  \begin{equation}
 \small
 \label{opt5}
 \begin{aligned}
      \mathcal{ P}_{2-2}:   {\boldsymbol  \theta}^{\ast}= & \arg\max\limits_{\boldsymbol  \theta}  \; \;\;\; \left\{d^{\boldsymbol {\mathcal X}  }_{\mathrm {MMPD}} ,   \tau_{\mathrm {min}}^{\boldsymbol {\mathcal X}  }  \right\} \\
     &    \text {Subject to}   \; \; \;  (\ref{opt2:a}).
     \end{aligned}
 \end{equation}

Unfortunately, there is no optimal solution for     (\ref{opt5}).  This is because  once the   sparse codebook ${\boldsymbol {\mathcal X}}_j$ with full DO and  optimal  MMPD is achieved,    $\mathbf{R}_j$ is determined, indicating there is no degree of freedom to improve  $\tau_{\mathrm {min}}^{\boldsymbol {\mathcal X}}$. However, as mentioned, it is ideal  to maximize the DO and MMPD first and then  improve    $\tau_{\mathrm {min}}^{\boldsymbol {\mathcal X}}$.   Hence, we consider a  feasible way by transforming  (\ref{opt5}) into a multi-stage design problem. Specifically, the optimal rotation angles  that achieve full DO and optimal MMPD of $\mathbf p$ are first obtained.  Then, the rest of angles are determined by maximizing   $\tau_{\mathrm {min}}^{\boldsymbol {\mathcal X}}$. The detailed design process is summarized in \textbf{Algorithm \ref{CBPEP}}.             
 
 \textit{3) Perform bit labeling:} Similar to the  AMI based labeling scheme,   bit labeling is performed  according to the  metric given in (\ref{label_pep}).

\section{Numerical results}

In this section,  we conduct numerical evaluations of the proposed SSD-SCMA  in both  uncoded and  coded systems with various codebooks. We first  evaluate the computational complexity of the proposed codebook design scheme in Subsection \ref{complexity}.  Then, Subsection  \ref{UPLB}   compares the proposed PEP upper bound  with the  exact ABER, and the AMI lower bound  with the exact AMI. Then, the proposed AMI-CBs and P-CBs are presented   in Subsection  \ref{CB_PRO}.  Accordingly, with the proposed codebooks, we evaluate the BER performance of    SSD-SCMA and  C-SCMA  in both uncoded and BICM-IDD systems in  Subsection  \ref{BER_sec}.  
 The indicator matrix that presented in Fig. \ref{factor} with $K=4$, $J=6$ and $\lambda = 150 \%$ is employed.  For comparison, we consider  the original GAM (OGAM) codebook \cite{mheich2018design},      StarQAM codebook \cite{yu2015optimized}, Chen's codebook \cite{chen2020design} \footnote{For $M=8$, the NS-QAM proposed in \cite{chen2020design} is employed.}, Huang's codebook \cite{huang2021downlink} and  Jiang's codebook  \cite{JiangLow}.   This is because  all these codebooks  are designed for downlink channels and  achieve  good BER  performance.


\subsection{Computational Complexity}
 \label{complexity}

 \textit{1) AMI-CB design:} As discussed in Subsection \ref{MI_sec}, the direct calculation of AMI, i.e., (\ref{AMI})  is intractable and often calculated by the tedious  Monte Carlo method.   In general, at least  $N_\text{s} = 10^3$ noise and channel samples are required to accurately estimate the AMI and its gradient at each iteration. The resultant  the computational complexity for AMI can  be approximated as $\mathcal O \left(N_\text{s} M^{2J} \right)$. In contrast, the computational complexity of the proposed lower bound can be approximated as $\mathcal O \left(  M^{2J} \right)$, which is significantly smaller than that of AMI. In addition, compared to the codebook design scheme proposed in \cite{JiangLow},  our proposed lower bound of AMI  in Rayleigh fading channels has closed-form expression and with simpler   computation  of $\text{Log}(\cdot)$ operations.

\textit{2) P-CB design:} The computational complexity of calculating the MMPD   is negligible and the main computational complexities of     {\textbf{Algorithm}}  \ref{CBPEP} are imposed by the  operations of permutation, labeling,  and the maximizing of   $\tau_{\mathrm {min}}^{\boldsymbol {\mathcal X}}$. With the aid of BSA algorithm, the computational complexity of  permutation and labeling can be reduced to  $\mathcal O \left( M^2\right)$. In addition, the complexity of determining the $\theta_3$ by maximizing     $\tau_{\mathrm {min}}^{\boldsymbol {\mathcal X}}$ can be approximated as  $\mathcal O \left( KM^{d_f}\right)$ \cite{chen2020design}. In general,  the design complexity of P-CB is relatively smaller than that of AMI-CB.


\subsection{Evaluations of PEP upper bound and AMI lower bound}
\label{UPLB}
\begin{figure*}[h]
\minipage{0.45\textwidth}
  \includegraphics[width=0.96 \textwidth]{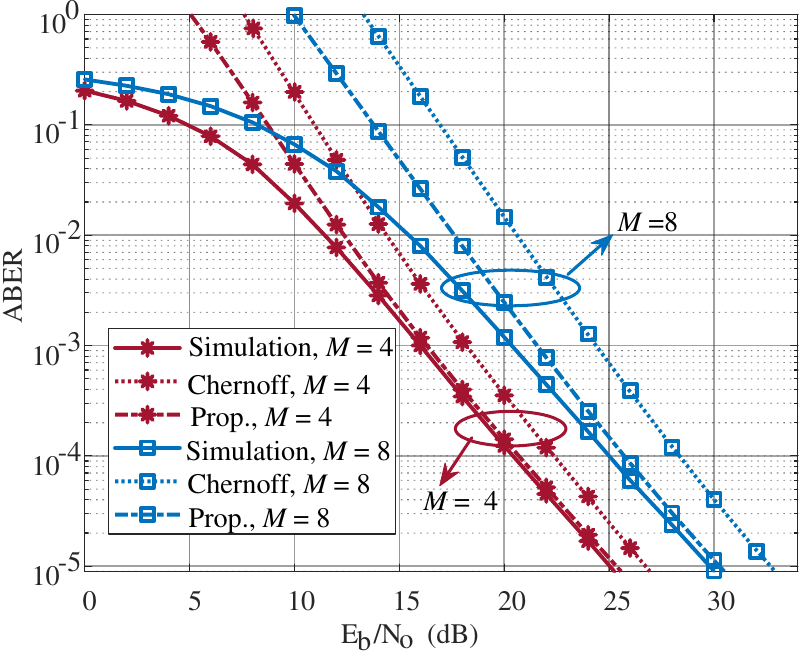}
  \caption{ABER against  various upper bounds for SSD-SCMA system.}\label{ABER}
\endminipage\hfill
\minipage{0.45 \textwidth}
  \includegraphics[width= 0.90 \textwidth]{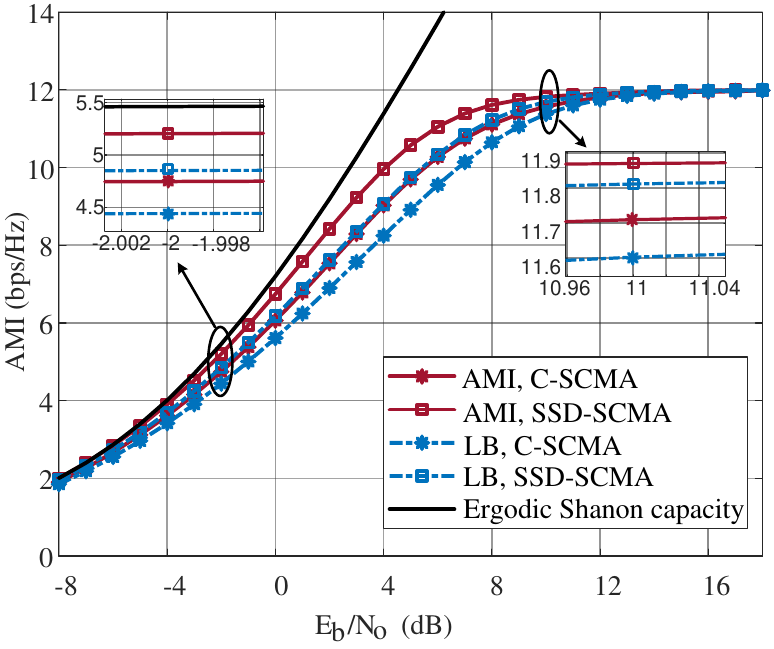}
  \caption{The AMI and the lower bound of AMI for SSD-SCMA and  C-SCMA systems.}\label{AMI_fig}
\endminipage\hfill
\end{figure*}

We now   compare the ABER of   Chernoff bound  with the   proposed PEP upper bound  in  (\ref{PEP_final}) and (\ref{ABER_}) by employing the StarQAM codebook in SSD-SCMA. As shown in Fig. \ref{ABER},  for $M=4$,  the proposed   upper bound, i.e., the line ``Prop.'',  close to the simulated ABER at high SNR,  while the Chernoff bound is about $2$ dB away from the simulated  ABER.  For $M=8$, the Chernoff bound is about $3$ dB away the simulated  ABER, however, the proposed PEP  upper bound still holds very tight to the  simulated  ABER in  high $\text{E}_\text{b}/\text{N}_{\text{o}}$ regime.

  \begin{figure}
  \begin{subfigure}[b]{0.49\linewidth}
    \centering
    \includegraphics[width=0.65\linewidth]{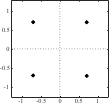} 
    \caption{QPSK, $d_{\min}= \sqrt{2}$} 
    \label{a} 
  \end{subfigure} 
  \begin{subfigure}[b]{0.49\linewidth}
    \centering
    \includegraphics[width=0.65\linewidth]{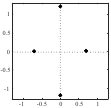} 
    \caption{$4$-TRI, $d_{\min}= \sqrt{2}$} 
    \label{b} 
  \end{subfigure}\\ 
  \begin{subfigure}[b]{0.49\linewidth}
    \centering
    \includegraphics[width=0.65\linewidth]{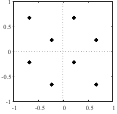} 
    \caption{$8$-NS-QAM,  $d_{\min}= 0.9$} 
    \label{c} 
  \end{subfigure}
  \begin{subfigure}[b]{0.5\linewidth}
    \centering
    \includegraphics[width=0.65\linewidth]{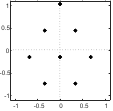} 
    \caption{$8$-TRI, $d_{\min}= 0.96$} 
    \label{d} 
  \end{subfigure} 
  \caption{ The adopted basic one dimensional constellation $\mathbf p$ with large MED ($d_{\min}$). The resultant codebooks are respectively denoted as P-CB1, P-CB2, P-CB3 and P-CB4. }
  \label{MCs} 
\end{figure}

 As discussed in Subsection  \ref{MI_sec},   direct calculation of AMI, i.e., (\ref{AMI})  is intractable and hence  Monte Carlo method is applied.   Suppose  about $N_\text{s} $ noise and channel samples are required to accurately estimate the AMI.  Hence the computational complexity can be approximated as $\mathcal O \left(N_\text{s } M^{2J} \right)$, which   may be not affordable   for  $M \geq 8$.  As mentioned in  \textit{Remark 2}, there exists a constant gap between the   AMI and the proposed lower bound and this can be leveraged for AMI analysis.   Fig. \ref{AMI_fig} shows the   AMI obtained by Monte Carlo simulation and the proposed AMI lower bound with a constant shift for both SSD-SCMA and  C-SCMA    with  $M=4$. One can see that the    proposed AMI lower bound plus a constant fits well the  simulated  AMI over the low and high SNR ranges.  For the other middle-range, there exists a small gap between the estimated  AMI and the AMI lower bound with a constant shift. However,  it is still effective to  improve the AMI  by maximizing  the lower bound.

 \begin{table} 
\small
\setlength{\tabcolsep}{1.6pt}
\begin{center}
\caption{\label{MPD_CB} The proposed P-CBs. }
\begin{tabular}{cccccc}
\toprule Codebook & \makecell[c] {Basic\\constellation}     &  $\left [\theta_{1},\theta_{2},\theta_{3} \right ]$  &  MMPD    \\
\midrule  P-CB1 &  QPSK     &           $\left [0.172 \pi, -0.172 \pi,  0.068 \pi\right ]$ & 0.86    \\
            P-CB2 & 4-TRI        &      $\left [0.083 \pi, -0.083 \pi, 0.336 \pi \right ]$ & 0.50 \\
          P-CB3 & 8-NS-QAM        &    $\left [0.072 \pi, -0.072 \pi, 0.125 \pi \right ]$& 0.35     \\
          P-CB4  &  8-TRI        &    $\left [ 0.075 \pi, -0.075 \pi, 0.378 \pi\right ]$ & 0.36   \\
\bottomrule
 \end{tabular}
\end{center}
\vspace{-1em}
 \end{table}
 
 \begin{table} 
\small
\begin{center}
\caption{\label{MIAll} A comparison   of $ \mathcal  I_{LB}^{\boldsymbol {\mathcal X}}$ with a constant shift at $\text{E}_{\text{b}} / \text{N}_{\text {o}} =5 $ dB and    $\text{E}_{\text{b}} / \text{N}_{\text {o}} =12 $ dB  for $M=4$ and $M=8$, respectively. }
\begin{tabular}{c|c|c|c}
   \hline
  $M$ &  Codebook &  C-SCMA &  SSD-SCMA    \\
     \hline
     \multirow{4}{*}{$M=4$} 
       &     AMI-CB1 &  $9.16$     &     $10.20$       \\
          \cline{2-4}
          &  P-CB1    &  $8.89$    &      $9.93$       \\
           \cline{2-4}
        &  Chen   \cite{chen2020design}   &   $8.89$     &    $9.85$       \\
               \cline{2-4}
        &  Jiang \cite{JiangLow}& $9.13$     &     $9.91$     \\    
           \hline  
       \multirow{4}{*}{$M=8$} 
       &     AMI-CB2 &  $14.82$     &     $15.54$       \\
          \cline{2-4}
          &  P-CB3    &  $14.60$    &      $15.18$       \\
           \cline{2-4}
        &  Chen   \cite{chen2020design}   &   $14.58$     &    $15.11$       \\
               \cline{2-4}
        &  Jiang  \cite{JiangLow}  &  $14.81$     &     $15.30$     \\    
           \hline           
 \end{tabular}
\end{center}
\vspace{-2em}
 \end{table}

 \subsection{The Proposed Codebooks}
 \label{CB_PRO}

\textit{1) The proposed P-CBs:}   We consider the following  one dimensional basic constellations $\mathbf p$ employed for the PEP based codebook: quadrature phase shift keying (QPSK), non-square quadrature amplitude modulation  (NS-QAM)  and the constellation drawn from a lattice of equilateral triangles \cite{Weifeng}, denoted as $M$-TRI.  The MED  of the basic constellation, denoted by $d_{\min}$, is also given.   
Based on  QPSK, $4$-TRI, NS-QAM and  $8$-TRI,  the resultant codebooks are respectively denoted as  P-CB1,  P-CB2,  P-CB3 and  P-CB4. The labeling     is carried out  at $\text{E}_{\text{b}} / \text{N}_{\text {o}} =15 $  dB for both $M=4$ and $M= 8$. 
The   rotation angles  and the MMPDs corresponding to $\theta_{\text{opt}}$  are also summarized in Table \ref{MPD_CB}. 





 \textit{2) The proposed AMI-CBs:} 
In  Algorithm \ref{CB_Mi}, we set   $I_1 =10$ and $I_2=25$ to generate the AMI-CBs. Table  \ref{MIAll} compares    the $\mathcal I_{LB}^{\boldsymbol {\mathcal X}}$     with a constant shift    of $-K \left( {1} / {\ln2}- 1 \right)$ for  various codebooks.     The proposed AMI-CBs own larger value of $ \mathcal 
 I_{LB}^{\boldsymbol {\mathcal X}}$, especially for SSD-SCMA system.  As discussed in  Section \ref{SymboticRela}, a codebook with larger MMPD (MPD) owns larger  $ \mathcal  I_{LB}^{\boldsymbol {\mathcal X}}$ of SSD-SCMA (conventional SCMA). We now give an example to illustrate this relationship. Consider the C-SCMA system, the MPDs for the proposed P-CB1, Chen's codebook and StarQAM codebook are respectively  given as $1$, $1$ and $0.72$, whereas the   $ \mathcal  I_{LB}^{\boldsymbol {\mathcal X}}$ with a constant shift  for those codebooks at  $\text{E}_{\text{b}} / \text{N}_{\text {o}} =12 $ dB are $11.89$, $11.89$ and $11.78$, respectively. However, this may not hold for low and mid range SNRs, which can been seen in Table \ref{MIAll}. In general,   the asymptotic relationship holds  when $\text{E}_{\text{b}} / \text{N}_{\text {o}} \geq 12 $ dB and $\text{E}_{\text{b}} / \text{N}_{\text {o}} \geq 16$ dB for $M=4$ and $M=8$, respectively. The optimized parameters $\left[\psi,\xi, \theta_{1},\theta_{2},\theta_{3} \right]$ for AMI-CB1 and  AMI-CB2 are given as $[-0.525, 0.659, -0.109, 0.272, 0.555]$ and $[-0.452, 0.700, -0.131, 0.152, 0.455]$, respectively. Interested readers can find the proposed AMI-CB1 in Appendix \ref{cb}, and  more relevant results at our GitHub project\footnote{\url{https://github.com/ethanlq/SCMA-codebook}}.

\subsection{BER performance of the proposed SSD-SCMA} 
 \label{BER_sec}

  \begin{figure*}
\minipage{0.45\textwidth}
  \includegraphics[width=1 \textwidth]{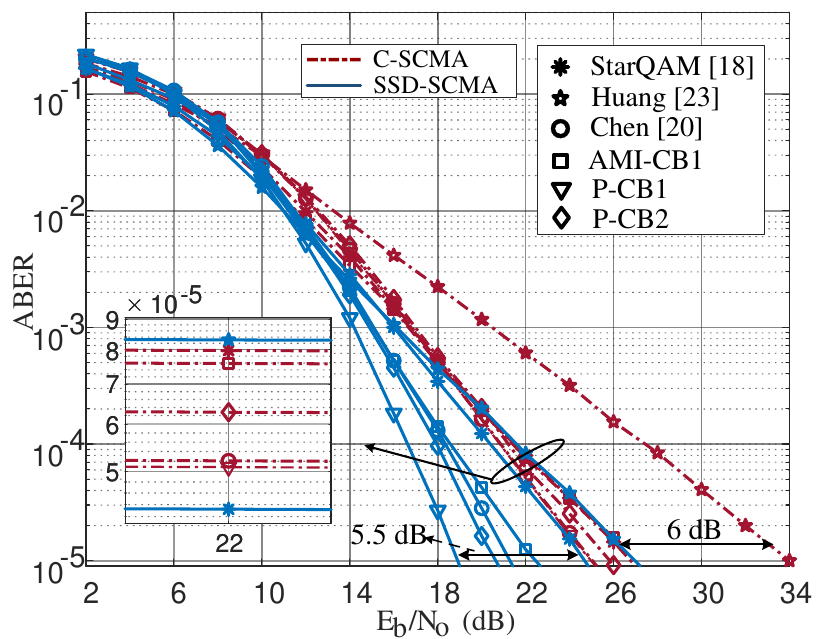}
  \caption{ABER comparison between SSD-SCMA and C-SCMA systems of various codebooks of $M=4$.}\label{BERM4}
\endminipage\hfill
\minipage{0.45 \textwidth}
  \includegraphics[width= 1 \textwidth]{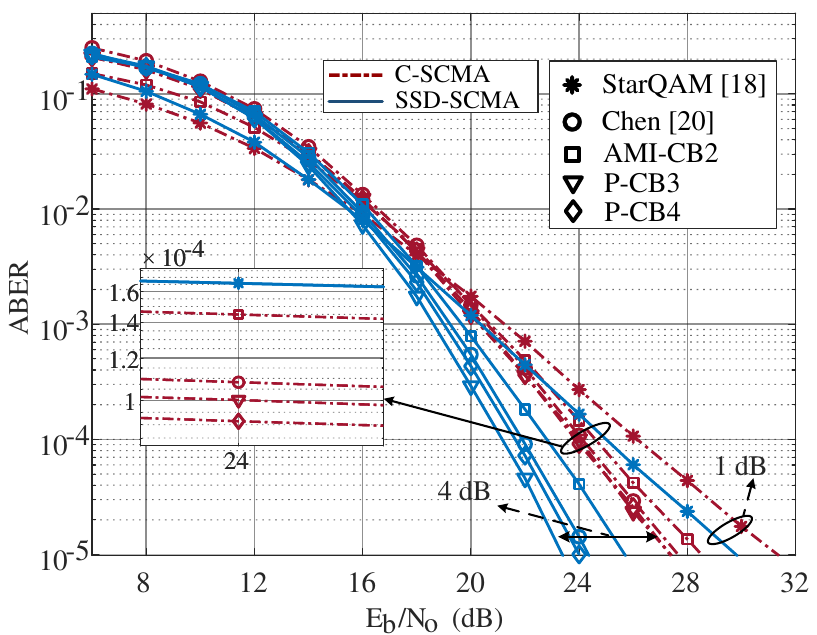}
  \caption{ABER comparison between SSD-SCMA and C-SCMA systems of various codebooks of $M=8$.}\label{BERM8}
\endminipage\hfill
\end{figure*}

 \textit{1) Uncoded system:} We first   evaluate the uncoded BER performance of   SSD-SCMA system with P-CBs, as shown in  Fig. \ref{BERM4}  and Fig. \ref{BERM8}. The dash lines denote the BERs for    conventional SCMA, whereas the solid lines are the proposed SSD-SCMA. The main observations are summarized as follows: 
\begin{itemize}
     \item The proposed SSD-SCMA    outperforms    C-SCMA   for both $M=4$ and $M=8$. For $M=4$,   $5.5$ dB  can be observed for SSD-SCMA with the proposed P-CB1 over that of  C-SCMA with Chen's codebook  at $\text{BER}= 10^{-5}$, whereas  $4$ dB gain is achieved for $M=8$ with the proposed P-CB3. 
  In addition,   SSD-SCMA   exhibits steeper BER slopes    than that of  C-SCMA  due to the larger DO of the former.  
  \item Not  all   codebooks can achieve such large gain for   SSD-SCMA system. For example, only $2$ dB gain is  observed for   Star-QAM codebook. Interestingly,  the proposed SSD-SCMA can improve the BER performance of Huang's codebook by $6$ dB. 
It is noted that the advantage of  Huang's codebook is in Gaussian channels, however, we show that its BER performance is significantly improved with the proposed SSD-SCMA over Rayleigh fading channels.

   \item The  proposed P-CB1 and PCB3 achieve  the best BER performance  for both SSD-SCMA  and C-SCMA systems, owning to the well optimized  MPD, MMPD, DO and labeling.

\end{itemize}

  \begin{figure*}
\minipage{0.45 \textwidth}
  \includegraphics[width=1.0 \linewidth]{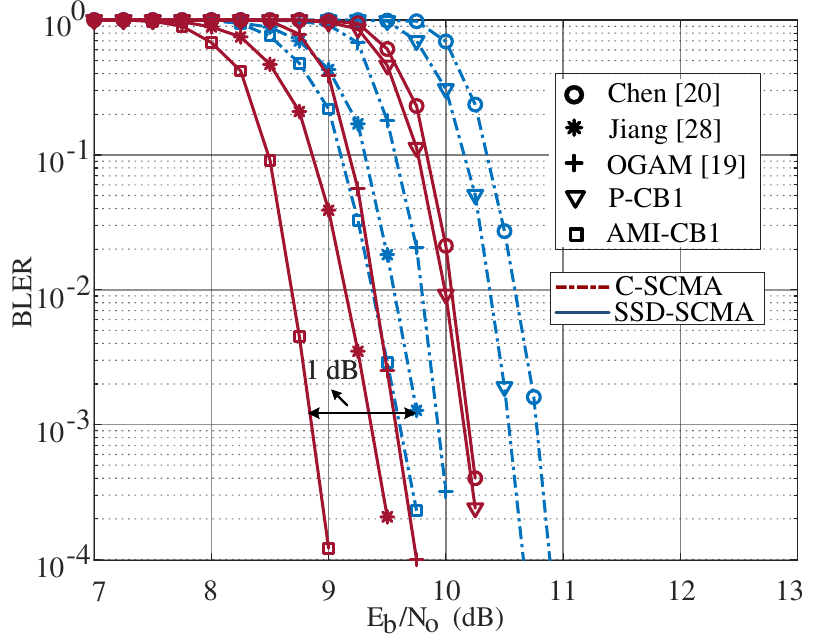}
  \caption{BLER comparison between SSD-SCMA and C-SCMA systems of various codebooks for $M=4$. }\label{LDPC_M4}
\endminipage\hfill
\minipage{0.45 \textwidth}
  \includegraphics[width=1.0 \linewidth]{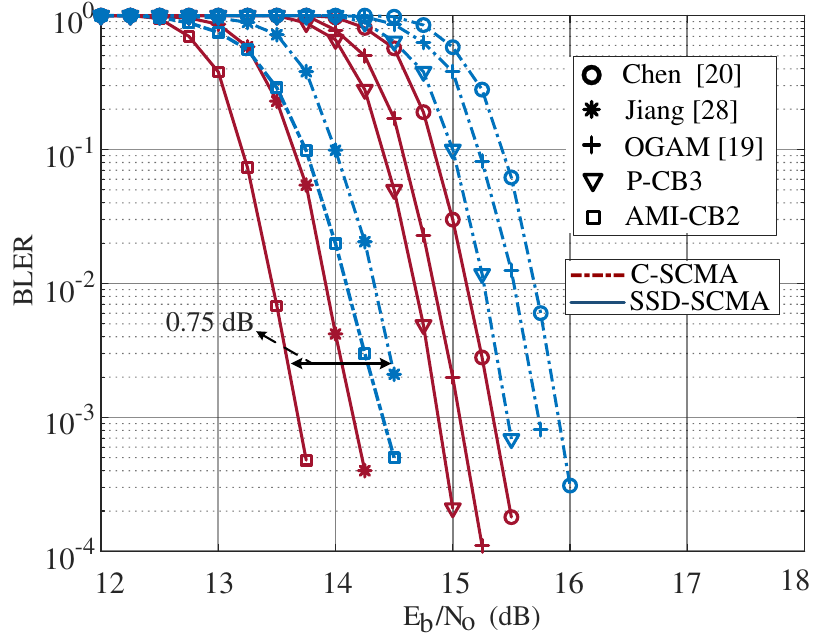}
  \caption{BLER comparison between SSD-SCMA and C-SCMA systems of various codebooks for $M=8$. }\label{LDPC_M8}
\endminipage\hfill
\end{figure*}

\textit{2) BICM-IDD system:}  Next, we evaluate the BER performance of coded    SSD-SCMA with   AMI-CBs under BICM-IDD receiver structure.  
Specifically,  the  5G NR LDPC code, as specified in TS38.212 \cite{5G_NR}, with block length of $1024$ and rate of $5/6$ are considered. The $\text{E}\text{b}/\text{N}{\text{o}}$ for  optimizing a codebook    is determined according to the code rate \cite{CodedXie}. Specifically, the $\text{E}\text{b}/\text{N}{\text{o}}$ that achieves an AMI of $rJ\log(M)$ is considered, where $r$ denotes the code rate. Namely, for a code rate of $5/6$, we employ the codebooks of $M=4$ and $M=8$ that are optimized at $\text{E}\text{b}/\text{N}{\text{o}} =4$ dB and $\text{E}\text{b}/\text{N}{\text{o}} =12$ dB, respectively.  The number of MPA    iterations is   $3$, the maximum number of LDPC decoding iterations is $25$ and the number of  BICM-IDD  iterations is $4$.

Fig.  \ref{LDPC_M4} and Fig.  \ref{LDPC_M8} show the block error rate (BLER) performances  of various codebooks for $M=4$ and $M=8$, respectively. The main observations are summarized as follows:
\begin{itemize}

\item The proposed SSD-SCMA  with the proposed AMI-CBs   achieve  the best BLER performance among all benchmarking  codebooks due  to its optimized AMI and labeling. Specifically, the proposed  SSD-SCMA with AMI-CB1 and  AMI-CB2 can achieve  about  $1$  dB and $0.75$ dB gains  compared to   C-SCMA    with the StarQAM codebook for $M=4$ and $M=8$, respectively.

\item  It is worth mentioning that a codebook that achieves better BER performance in an uncoded system may not outperform    in a BICM-IDD system, and vice versa. For example, the proposed PCB1 and Chen' codebook achieve better BER performance than the Star-QAM, OGAM codebooks and the proposed AMI-CB1 in    uncoded C-SCMA  with $M=4$; however, their error performances deteriorate under the BICM-IDD system.   This is because their    performance metrics and codebook design criteria are different from each other.
\end{itemize}

\section{Conclusions}
In this paper,  we  have  introduced a novel  SSD-SCMA    that  can significantly improve the error performance  of SCMA in downlink Rayleigh fading channels.
We  have  analyzed the   AMI and PEP  of SSD-SCMA,    and derived the AMI lower bound and PEP upper bound.    Based on the proposed bounds, a systematic  codebook  design metrics  have been established for both AMI-CBs and P-CBs. The asymptotic relationship between the two codebook designs based on PEP and AMI   have been revealed.   Furthermore, a novel  E-GAM has been proposed as the MC to design the AMI-CBs, whereas an efficient approach by permuting a basic constellation has been introduced to design   P-CBs. 
Numerical results have demonstrated the  advantages  of the proposed SSD-SCMA with the proposed AMI-CBs and P-CBs in both uncoded and   BICM-IDD systems. 

  \appendices
 
  \section{The proposed  codebooks} 
  \label{cb}
 
The proposed  AMI-CB1   is presented, and the other codebooks can be constructed with the presented parameters. Alternatively, one can find these codebooks at:   \url{https://github.com/ethanlq/SCMA-codebook}. 
  For  $M=4$, the four columns of   $\mathcal X_j, j=1,2, \ldots, J$  denote the codewords   labelled by $00$, $01$, $10$, $11$, respectively. 
         \vspace{-1em} 
      \begin{equation}
  \setlength{\arraycolsep}{2.0pt} 
  \footnotesize   
     \mathcal X_1=
\left[\begin{matrix}
   0&   0.0432 - 0.7904i&   0&   -0.5417 - 0.2827i\\
   0&   -0.1541 - 0.3106i&   0&   0.5097 + 0.7873i\\
   0&   0.1541 + 0.3106i&   0&   -0.5097 - 0.7873i\\
   0&   -0.0432 + 0.7904i&   0&   0.5417 + 0.2827i\\
      \end{matrix}\right]^{\mathcal{T}},  \notag
        \end{equation}  
     \vspace{-0.8em}     
      \begin{equation}
  \setlength{\arraycolsep}{2.0pt} 
  \footnotesize   
     \mathcal X_2=
\left[\begin{matrix}
   0.3884 - 0.4718i&   0&   0.0432 - 0.7904i&   0\\
  -0.8755 + 0.3364i&   0&   -0.1541 - 0.3106i&   0\\
   0.8755 - 0.3364i&   0&   0.1541 + 0.3106i&   0\\
  -0.3884 + 0.4718i&   0&   -0.0432 + 0.7904i&   0\\
      \end{matrix}\right]^{\mathcal{T}},  \notag
        \end{equation}  
     \vspace{-0.8em}     
      \begin{equation}
  \setlength{\arraycolsep}{2.0pt} 
  \footnotesize   
     \mathcal X_3=
\left[\begin{matrix}
  -0.5853 - 0.5329i&   0.3884 - 0.4718i&   0&   0\\
  -0.3382 - 0.0768i&   -0.8755 + 0.3364i&   0&   0\\
   0.3382 + 0.0768i&   0.8755 - 0.3364i&   0&   0\\
   0.5853 + 0.5329i&   -0.3884 + 0.4718i&   0&   0\\
      \end{matrix}\right]^{\mathcal{T}},  \notag
        \end{equation}  
     \vspace{-0.8em}     
      \begin{equation}
  \setlength{\arraycolsep}{2.0pt} 
  \footnotesize   
     \mathcal X_4=
\left[\begin{matrix}
   0&   0&   -0.5417 - 0.2827i&   0.0432 - 0.7904i\\
   0&   0&   0.5097 + 0.7873i&   -0.1541 - 0.3106i\\
   0&   0&   -0.5097 - 0.7873i&   0.1541 + 0.3106i\\
   0&   0&   0.5417 + 0.2827i&   -0.0432 + 0.7904i\\
      \end{matrix}\right]^{\mathcal{T}},  \notag
        \end{equation}  
     \vspace{-0.8em}     
      \begin{equation}
  \setlength{\arraycolsep}{2.0pt} 
  \footnotesize   
     \mathcal X_5=
\left[\begin{matrix}
   0.0432 - 0.7904i&   0&   0&   0.3884 - 0.4718i\\
  -0.1541 - 0.3106i&   0&   0&   -0.8755 + 0.3364i\\
   0.1541 + 0.3106i&   0&   0&   0.8755 - 0.3364i\\
  -0.0432 + 0.7904i&   0&   0&   -0.3884 + 0.4718i\\
      \end{matrix}\right]^{\mathcal{T}},  \notag
        \end{equation}  
     \vspace{-0.8em}     
      \begin{equation}
  \setlength{\arraycolsep}{2.0pt} 
  \footnotesize   
     \mathcal X_6=
\left[\begin{matrix}
   0&   -0.5417 - 0.2827i&   0.6422 - 0.4628i&   0\\
   0&   0.5097 + 0.7873i&   0.1449 - 0.3150i&   0\\
   0&   -0.5097 - 0.7873i&   -0.1449 + 0.3150i&   0\\
   0&   0.5417 + 0.2827i&   -0.6422 + 0.4628i&   0\\
      \end{matrix}\right]^{\mathcal{T}}. \notag    
        \end{equation}

\bibliography{ref} 
\bibliographystyle{IEEEtran}

\end{document}